\documentclass[aps,pra,twocolumn,showpacs,preprintnumbers,
               superscriptaddress, nofootinbib,float]{revtex4-1}
\usepackage{graphicx, fancybox}
\usepackage{amsmath,amssymb}
\usepackage[colorlinks=true, pdfstartview=FitV, linkcolor=red,
citecolor=blue, urlcolor=blue]{hyperref}
\usepackage{color}
\usepackage{soul}

\newcommand{\comment}[1]{}

\newcommand{\be}{\begin{equation}}
\newcommand{\ee}{\end{equation}}
\newcommand{\ba}{\begin{eqnarray}}
\newcommand{\ea}{\end{eqnarray}}

\newcommand{\gsim}{\mathrel{\hbox{\rlap{\lower.55ex \hbox {$\sim$}}
                   \kern-.3em \raise.4ex \hbox{$>$}}}}
\newcommand{\lsim}{\mathrel{\hbox{\rlap{\lower.55ex \hbox {$\sim$}}
                   \kern-.3em \raise.4ex \hbox{$<$}}}}

\def\be{\begin{eqnarray}}
\def\ee{\end{eqnarray}}

\def\roughly#1{\mathrel{\raise.3ex\hbox{$#1$\kern-.75em%
\lower1ex\hbox{$\sim$}}}}
\def\lsim{\roughly<}
\def\gsim{\roughly>}

\def\({\left(}
\def\){\right)}
\def\[{\left[}
\def\]{\right]}

\def\lsim{\mathrel{\rlap{\lower3pt\hbox{\hskip1pt$\sim$}}
     \raise1pt\hbox{$<$}}} 
\def\gsim{\mathrel{\rlap{\lower3pt\hbox{\hskip1pt$\sim$}}
     \raise1pt\hbox{$>$}}} 

\usepackage[normalem]{ulem}

\begin{document}

\preprint{BNL-105041-2014-JA, RBRC-1093, RIKEN-QHP-151}

\title{Production and Elliptic Flow of Dileptons and Photons
in the semi-Quark Gluon Plasma}

\author{Charles Gale}
\affiliation{Department of Physics, McGill University,
3600 University Street, Montreal, QC H3A 2T8, Canada}
\affiliation{Frankfurt Institute for Advanced Studies, Ruth-Moufang-Str. 1, D-60438 Frankfurt am Main, Germany}

\author{Yoshimasa Hidaka}
\affiliation{Theoretical Research Division,
Nishina Center, RIKEN, Wako 351-0198, Japan}

\author{Sangyong Jeon}
\affiliation{Department of Physics, McGill University,
3600 University Street, Montreal, QC H3A 2T8, Canada}

\author{Shu Lin}
\affiliation{RIKEN/BNL Research Center,
Brookhaven National Laboratory, Upton, NY 11973, USA}

\author{Jean-Fran\c{c}ois Paquet}
\affiliation{Department of Physics, McGill University,
3600 University Street, Montreal, QC H3A 2T8, Canada}

\author{Robert D. Pisarski}
\affiliation{Department of Physics,
Brookhaven National Laboratory, Upton, NY 11973, USA}
\affiliation{RIKEN/BNL Research Center,
Brookhaven National Laboratory, Upton, NY 11973, USA}

\author{Daisuke Satow}
\affiliation{Theoretical Research Division,
Nishina Center, RIKEN, Wako 351-0198, Japan}
\affiliation{Department of Physics,
Brookhaven National Laboratory, Upton, NY 11973, USA}

\author{Vladimir V. Skokov}
\affiliation{Department of Physics,
Western Michigan University, 1903 W. Michigan Avenue, Kalamazoo, MI 49008}

\author{Gojko Vujanovic}
\affiliation{Department of Physics, McGill University,
3600 University Street, Montreal, QC H3A 2T8, Canada}

\begin{abstract}
We consider the thermal production of 
dileptons and photons at temperatures above the
critical temperature in QCD.  We use a model where
color excitations are suppressed by a small value of the Polyakov loop,
the semi Quark-Gluon Plasma (QGP).
Comparing the semi-QGP to the perturbative QGP, we find
a mild enhancement of thermal dileptons.
In contrast, to leading logarithmic order in weak coupling
there are far fewer hard photons from the semi-QGP than the usual QGP.
To illustrate the possible effects 
on photon and dileptons production in heavy ion collisions, 
we integrate the rate with a realistic hydrodynamic simulation.
Dileptons uniformly exhibit a small flow, 
but the strong suppression of photons in the semi-QGP tends to bias
the elliptical flow of photons to that generated in the hadronic phase.
\end{abstract}
\maketitle

The collisions of heavy nuclei at ultra-relativistic energies
can be used to investigate 
the properties of the Quark-Gluon Plasma (QGP).
At both
the Relativistic Heavy Ion Collider (RHIC) and the Large Hadron Collider (LHC),
much of the collision takes place at temperatures which are not that far
above that for the transition, $T_c$.
This is a difficult region to study: perturbative methods can be used
at high temperature, but not near $T_c$ \cite{Haque:2014rua}.
Similarly, hadronic models
are valid at low temperature, but break down near $T_c$
\cite{Huovinen:2009yb, *Andronic:2012ut}.
One model of the region above but near $T_c$ is the semi-QGP
\cite{Pisarski:2006hz, Hidaka:2008dr, *Hidaka:2009hs, *Hidaka:2009xh,
*Hidaka:2009ma, Dumitru:2010mj, *Dumitru:2012fw, *Kashiwa:2012wa,
*Pisarski:2012bj, *Kashiwa:2013rm, *Lin:2013qu, Lin:2013efa}.
This incorporates the results of numerical simulations on the lattice
\cite{Bazavov:2009zn, *DeTar:2009ef, *Fodor:2009ax, *Petreczky:2012rq,
*Borsanyi:2013bia, *Bhattacharya:2014ara, *Sharma:2013hsa},
which show that colored excitations are strongly suppressed
when $T \rightarrow T_c^+$, as
the expectation value of the Polyakov loop decreases markedly.

A notable property of heavy ion collisions is elliptic flow, how the
initial spatial anisotropy of peripheral collisions is transformed into
a momentum anisotropy.  
The large
elliptic flow of hadrons can be well modeled by hydrodynamic models
in which the QCD medium is close to an ideal fluid 
\cite{Heinz:2009xj, *Heinz:2013th, *Gale:2013da, Schenke:2010nt, Schenke:2010rr}.

Electromagnetic signals, such as dilepton or photon production, are
another valuable probe,
since they reflect properties of the quark and gluon
distributions of the QGP, and once produced, escape without
significant interaction
\cite{Baier:1991em, *Kapusta:1991qp, Steele:1996su, *Steele:1997tv, *Steele:1999hf, *Dusling:2006yv, *Dusling:2007su, *Dusling:2009ej, Lee:1998nz, Aurenche:2000gf, Arnold:2002ja, *Arnold:2001ms, *Arnold:2002ja, Turbide:2003si, *Manninen:2010yf, *Staig:2010by, *Linnyk:2011vx, *Linnyk:2012pu, *Rapp:2013nxa, *Hohler:2013eba, *Lee:2014pwa, Vujanovic:2013jpa, Chatterjee:2005de, *Bratkovskaya:2008iq, *vanHees:2011vb, *Basar:2012bp, *Bzdak:2012fr, *Fukushima:2012fg, *Liu:2012ax, *Shen:2013cca, *Shen:2013vja, *Linnyk:2013hta, *Linnyk:2013wma, *Muller:2013ila, *Basar:2014swa,*vanHees:2014ida, *Monnai:2014kqa, McLerran:2014hza, Dion:2011pp, Klasen:2013mga, *Klasen:2014xfa, Arleo:2006xb, *Arleo:2008dn, *Arleo:2011gc, Adare:2011zr, Lohner:2012ct, Sakaguchi:2014ewa}.  
For example, if most photons are emitted at
high temperature in the QGP, since the flow at early times is small, one
would expect a small net elliptic flow for photons.
However, recently both the PHENIX experiment at RHIC \cite{Adare:2011zr}
and the ALICE experiment at the LHC \cite{Lohner:2012ct}
have found a large elliptic flow for photons, comparable to that of
hadrons.  This is most puzzling
\cite{Chatterjee:2005de, *Bratkovskaya:2008iq, *vanHees:2011vb, *Basar:2012bp, *Bzdak:2012fr, *Fukushima:2012fg, *Liu:2012ax, *Shen:2013cca, *Shen:2013vja, *Linnyk:2013hta, *Linnyk:2013wma, *Muller:2013ila, *Basar:2014swa,*vanHees:2014ida, *Monnai:2014kqa, McLerran:2014hza, Sakaguchi:2014ewa}.  

In this paper we present the results
for the thermal production of hard dileptons and photons in
the semi-QGP, and compare them with those of the perturbative QGP.  
Surprisingly, we find a sharp qualitative difference
between the two.  
In the semi-QGP, the production of dileptons is similar between the
deconfined and confined phases, while photon production is
{\it strongly} suppressed near $T_c$.
We compute to leading order in the QCD coupling 
(for photons, only to leading logarithmic order)
and give complete results later \cite{Future}.
We then use a hydrodynamic model \cite{Vujanovic:2013jpa} to compute
the effect on the number of dileptons and photons produced, and on
the elliptic flow, $v_2$.  
The effects on thermal dileptons are modest.
The suppression of thermal photons near $T_c$ in the semi-QGP, though,
implies that $v_2$ is biased 
towards that generated in the hadronic phase.
Our results may help to understand the puzzle
of the elliptic flow for photons.

Deconfinement in a $SU(N_c)$ gauge theory
is characterized by the Polyakov loop,
$\ell = (1/N_c) {{\rm tr}} \, {\cal P} \exp(i g\int^{1/T}_0 d\tau \, A_0)
$, where $\cal P$ denotes path ordering,
$T$ is the temperature, $g$ the gauge coupling constant,
and $A_0$ the temporal component of the gauge field.
At high temperature $\langle \ell \rangle \sim 1$ 
~\cite{Gava:1981qd, *Burnier:2009bk, *Brambilla:2010xn}, while
$\langle \ell \rangle = 0$ in the confined
phase of a pure gauge theory.
With dynamical quarks, $\langle \ell \rangle > 0$ at any nonzero
temperature, but lattice simulations show that the value of the
(renormalized) loop is small at
$T_c$, $\langle \ell(T_c) \rangle \approx 0.1$
\cite{Bazavov:2009zn, DeTar:2009ef, *Fodor:2009ax, *Petreczky:2012rq,
*Borsanyi:2013bia, *Bhattacharya:2014ara, *Sharma:2013hsa}.  

The simplest way to represent a phase where $\langle \ell \rangle < 1$
is to work in mean field theory,
taking $A_0$ to be a constant, diagonal matrix, 
$(A_0^{cl})^{a b} = \delta^{a b} Q^a/g$
\cite{Pisarski:2006hz, Hidaka:2008dr, *Hidaka:2009hs, *Hidaka:2009xh,
*Hidaka:2009ma, Dumitru:2010mj, *Dumitru:2012fw, *Kashiwa:2012wa,
*Pisarski:2012bj, *Kashiwa:2013rm, *Lin:2013qu, Lin:2013efa}.
The Polyakov loop is then
$
\ell = 1/N_c \sum_a {\rm e}^{i Q^a/T} 
$ , where the color index $a = 1 \ldots N_c$.
For three colors, $A_0^{cl} = (Q,-Q,0)/g$, so
$Q = 2 \pi T/3$ in the confined vacuum, $\ell = 0$.
Since $A_0^{cl} \sim T/g$, this is manifestly a model
of non-perturbative physics.

In Minkowski spacetime, the diagrams are those of ordinary perturbation
theory, except that the 
background field $A_0^{cl}$ acts like an imaginary chemical potential
for color.
For a quark with color $a$, the Fermi-Dirac distribution
function is $1/(e^{(E - i Q^a)/T} + 1)$.
In the double line basis gluons carry two color indices, $(ab)$,
and their Bose-Einstein distribution function
involves a difference of $Q$'s, $1/(e^{(E - i (Q^a-Q^b))/T} - 1)$.  
In the Boltzmann approximation, the distribution function for a single
quark (or anti-quark), summed over color, is 
suppressed by the Polyakov loop, $\sim \sum_a e^{-(E -i Q^a)/T}/N_c \sim 
e^{-E/T} \ell$; for gluons, it is $\sim e^{-E/T} \ell^2$.

In the semi-QGP model, one computes to leading order in the QCD
coupling with $Q^a \neq 0$
\cite{Pisarski:2006hz, Hidaka:2008dr, *Hidaka:2009hs, *Hidaka:2009xh,
*Hidaka:2009ma, Dumitru:2010mj, *Dumitru:2012fw, *Kashiwa:2012wa,
*Pisarski:2012bj, *Kashiwa:2013rm, *Lin:2013qu, Lin:2013efa}.
We first discuss the results for thermal dilepton production.
Let the sum of the momenta of the dilepton be $P^\mu=
( E,\vec{p}\,)$, $p=|\vec{p}\,|$, where $E > p$.  
To leading order in perturbation theory, 
this arises from the annihilation of a quark anti-quark pair into a virtual
photon, which then decays into a dilepton pair.
For three colors and $Q=0$, the production rate~\cite{Vujanovic:2013jpa} is
\begin{align}
\label{eq:dilepton-result-Q=0}
\begin{split}
\left.  \frac{d\varGamma}{d^4P}\right |_{Q=0}
&= \frac{\alpha_{em}}{6\pi^4}\; n(E)
\left(1-\frac{2T}{p}
\ln\frac{1 + {\rm e }^{-p_-/T}}{1 + {\rm e }^{-p_+/T}}\right) ;
\end{split}
\end{align}
$p_\pm = (E \pm p)/2$, and
$n(E) = 1/({\rm e}^{E/T} - 1)$ is the usual Bose-Einstein distribution
function.  This includes
the contributions of (massless) up, down and strange quarks,
where $\alpha_{em} = e^2/4\pi$, and $e$ is 
the electromagnetic coupling constant.

In the semi-QGP, to leading order the result when $Q\neq 0$ is a simple
factor times that for $Q=0$ \cite{Future},
\begin{equation}
\left. \frac{d\varGamma}{d^4P}\right|_{Q\neq0}
= \left. f_{l\overline{l}}(Q)
\; \frac{d\varGamma}{d^4P}\right |_{Q=0} \; ,
\label{eq:dGamma}
\end{equation}
where $f_{l\overline{l}}(Q)\equiv 
\widetilde{f}_{l\overline{l}}(Q)/\widetilde{f}_{l\overline{l}}(0)$.
For three colors, this can be written 
in terms of the Polyakov loop,
\begin{equation}
\label{eq:dilepton-f-result-N=3}
\widetilde{f}_{l\overline{l}}=  1-\frac{2T}{3 p}
\ln\frac{1+3\ell {\rm e}^{-p_-/T}+3\ell {\rm e}^{-2p_-/T}+
{\rm e}^{-3 p_-/T}}
{1+3\ell \, {\rm e}^{-p_+/T}+
3 \ell {\rm e}^{-2 p_+/T}+ {\rm e}^{-3 p_+/T}} \; .
\end{equation}

\begin{figure}[t]
\begin{center}
\includegraphics[width=0.45\textwidth]{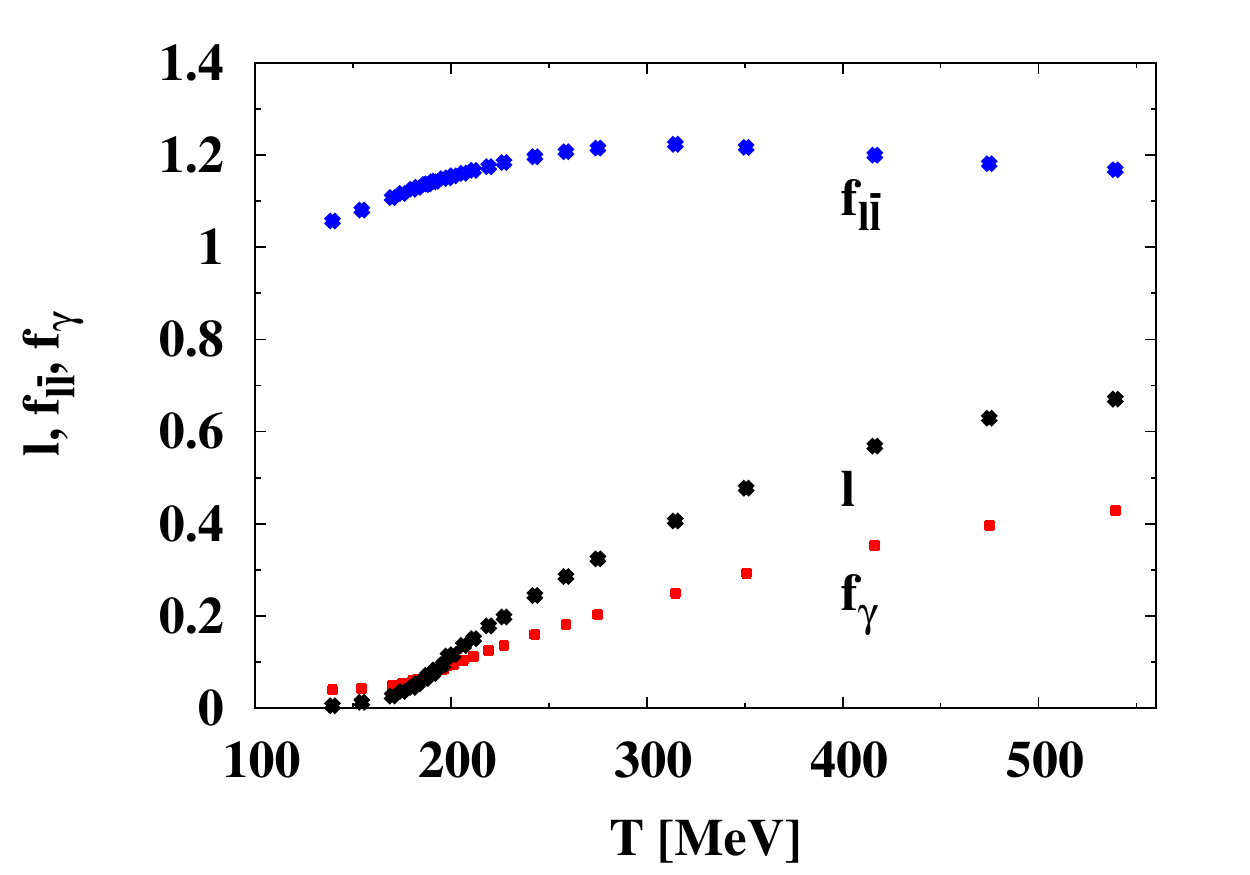}
\caption{
The ratio of the thermal production of dileptons and photons in the semi-QGP,
versus that in perturbation theory, as a function of temperature.
For dileptons, $f_{l\overline{l}}$ from
Eq.~(\ref{eq:dilepton-f-result-N=3}) is for
$E = 1$~GeV and $p=0$.
For photons, $f_\gamma$ in Eq.~(\ref{f_gamma}) is independent of the photon
momentum.  The loop is taken from Ref. ~\cite{Lin:2013efa}.
}
\label{fig:di_real}
\end{center}
\end{figure}

In the special case that the dileptons move back to back, $p=0$,
we plot the modification factor at $E = 1$~GeV
as a function of temperature in Fig.~(\ref{fig:di_real}),
taking the $Q^a$'s from Ref. \cite{Lin:2013efa}.
We find that 
$f_{l\overline{l}}(Q)$ is always {\it greater} than one.

To understand this, remember that in kinetic theory
the production rate for dileptons
is the product of statistical distribution functions
times an amplitude.  When $p = 0$,
the distribution functions are for a quark with energy $E/2$
and color $a$, and
an anti-quark, also with energy $E/2$ and color $a$.
If the total energy $E \gg T$,
we can use the Boltzmann approximation for the 
$Q^a$-dependent Fermi-Dirac distribution functions,
\begin{align}
e^2 \; \sum_{a=1}^{N} e^{-(E/2-iQ^a)/T}e^{-(E/2+iQ^a)/T}
\left|{\cal M}_{l\overline{l}}\right|^2 \; .
\label{dilepton_kinetic}
\end{align}
As the $Q^a$'s are like a chemical potential for color, they enter the
distribution functions for the quark and anti-quark with {\it opposite} signs,
and so at large energy, cancel identically.  
That is, the probability for a hard virtual photon to produce 
a quark anti-quark pair is independent of the $Q^a$'s, and so the Polyakov 
loop.  
This is in stark contrast to the statistical distribution function
for a {\it single} quark or anti-quark, which is $\sim \ell$.

Fig. (\ref{fig:di_real}) shows that 
for moderate values of $E \sim T$, there 
are corrections to the Boltzmann approximation
which even give a modest {\it enhancement} 
above $T_c$, by about $\sim 20\%$.

Expanding Eq. (\ref{eq:dilepton-f-result-N=3})
to quadratic order in the $Q^a$ is equivalent to considering
a condensate $\sim \langle {\rm tr}\, A_0^2 \rangle$, and agrees
with previous results \cite{Lee:1998nz}.  
Ref. \cite{Lee:1998nz} suggested that an enhancement like that which we
find could explain the excess of dileptons found below the $\rho$ meson mass
in heavy ion collisions; see, also, Ref.
\cite{Steele:1996su, *Steele:1997tv, *Steele:1999hf, *Dusling:2006yv, *Dusling:2007su, *Dusling:2009ej, Manninen:2010yf, *Staig:2010by, *Linnyk:2011vx, *Linnyk:2012pu, *Rapp:2013nxa, *Hohler:2013eba, *Lee:2014pwa}.

We now consider the production of real photons at a large momentum $P^\mu$, 
where $E= p \gg T$.
To leading order in the QCD coupling, 
apparently two processes contribute to photon production:
Compton scattering of a quark or anti-quark,
and the pair annihilation of a quark and an anti-quark.
These $2 \rightarrow 2$ processes 
\cite{Baier:1991em, *Kapusta:1991qp} 
are both $\sim e^2 g^2$.
However, a quark which 
scatters with an arbitrary number of soft gluons, with $E_{\rm soft}
\sim gT $, emits collinear photons at the same order, $\sim e^2 g^2$.
\cite{Aurenche:2000gf, Arnold:2002ja, *Arnold:2001ms, *Arnold:2002ja}. 
This depends crucially upon Bose-Einstein enhancement for
the soft gluon, as $n(E_{{\rm soft}}) \sim 1/g$.

In the semi-QGP, however, 
there is no Bose-Einstein enhancement for off-diagonal gluons:
at small $E$ the gluon distribution function is
$\sim 1/(e^{- i(Q^a - Q^b)/T} - 1)$, if $a \neq b$ and $Q^a - Q^b \sim T$.  
There is Bose-Einstein enhancement for soft, 
diagonal gluons, where $a = b$, but
at large $N_c$ there are only $\sim N_c$ diagonal gluons
to $\sim N_c^2$ off-diagonal gluons. 
Consequently up to corrections
$\sim 1/N_c$, in the semi-QGP the production of real photons is dominated by
$2 \rightarrow 2$ processes.  This is a straightforward generalization of
the original computations of Ref.~\cite{Baier:1991em, *Kapusta:1991qp}. 
The results for collinear emission at large $N_c$ will be given 
later~\cite{Future}.

Computing thermal photon production only to
leading logarithmic order, we find \cite{Future}
\begin{equation}
\left. E \frac{d \varGamma}{d^3p}\right|_{Q \neq 0}
=  f_\gamma(Q) \left. E \frac{d \varGamma}{d^3p}\right|_{Q=0} \; .
\label{define_f_gamma}
\end{equation}
At the same order, 
the result for $2 \rightarrow 2$ scattering 
in the perturbative regime~\cite{Baier:1991em, *Kapusta:1991qp} is
\begin{equation}
\left. E \frac{d \varGamma}{d^3p}\right|_{Q=0}
=  \frac{\alpha_{em} \alpha_s}{3 \pi^2} \; 
 {\rm e}^{-E/T} \; T^2 \; \ln \left( \frac{E}{g^2 T} \right) \; ,
\end{equation}
where $\alpha_s = g^2/(4 \pi)$, and
\begin{equation}\label{f_gamma}
f_\gamma(Q) = 1 - 4 \, q + \frac{10}{3} \, q^2 \;\; ; \;\;
q = \frac{Q}{2 \pi T}\;,\; 0<q<1 .
\end{equation}
In the perturbative limit, $f_\gamma(0) = 1$.  This function decreases
monotonically as $Q$ increases, with $f_\gamma(2 \pi T/3) = 1/27$
in the confined phase.
In Fig. ~(\ref{fig:di_real}) we plot $f_\gamma$ versus temperature.  This
result is independent of momentum when $E \gg T$.

Why photon production is strongly suppressed in the confined phase can be
understood from the case of pair annihilation.  
Using kinetic theory in the Boltzmann
approximation, photon production is proportional to
\begin{equation}
\label{kinetic}
e^2 \; g^2 \; \sum_{a,b}e^{-(E_1-iQ^a)/T}e^{-(E_2+iQ^b)/T}
|{\cal M}^{ab}_\gamma|^2
\; ,
\end{equation}
where $E_1$ is the energy of the incoming quark with color $a$, 
$E_2$ the energy of the anti-quark with color $b$,
and ${\cal M}^{ab}_\gamma$ a matrix element, which depends upon $a$ and $b$.
The quark and anti-quark
then scatter into a gluon, with color indices $(ab)$, 
and a photon.  In the deconfined
phase, the rate is $\sim e^2 g^2 N_c^2$.
In the confined phase, however, to avoid suppression by powers of the
Polyakov loop the color charges of the quark and anti-quark must
match up, with $a=b$.  This reduces the result by 
one factor of $1/N_c$.
Further, the matrix element 
${\cal M}^{ab}_\gamma$ involves the quark-gluon vertex;
when $a = b$, this gives another factor of $1/N_c$, for an 
overall factor of $1/N_c^2$.   
The same counting in $1/N_c$ applies for Compton scattering.  In all, 
at large $N_c$ the ratio of hard photon production in the confined phase,
to that in the deconfined phase, is $f_\gamma = 1/(3 N_c^2)$
\cite{Future}.  Even for three colors this is a 
rather small number, $1/27$.

\begin{figure}[t]
\begin{center}
\vspace{0.4cm}
\begin{tabular} {c c}
\includegraphics[width=0.36\textwidth]{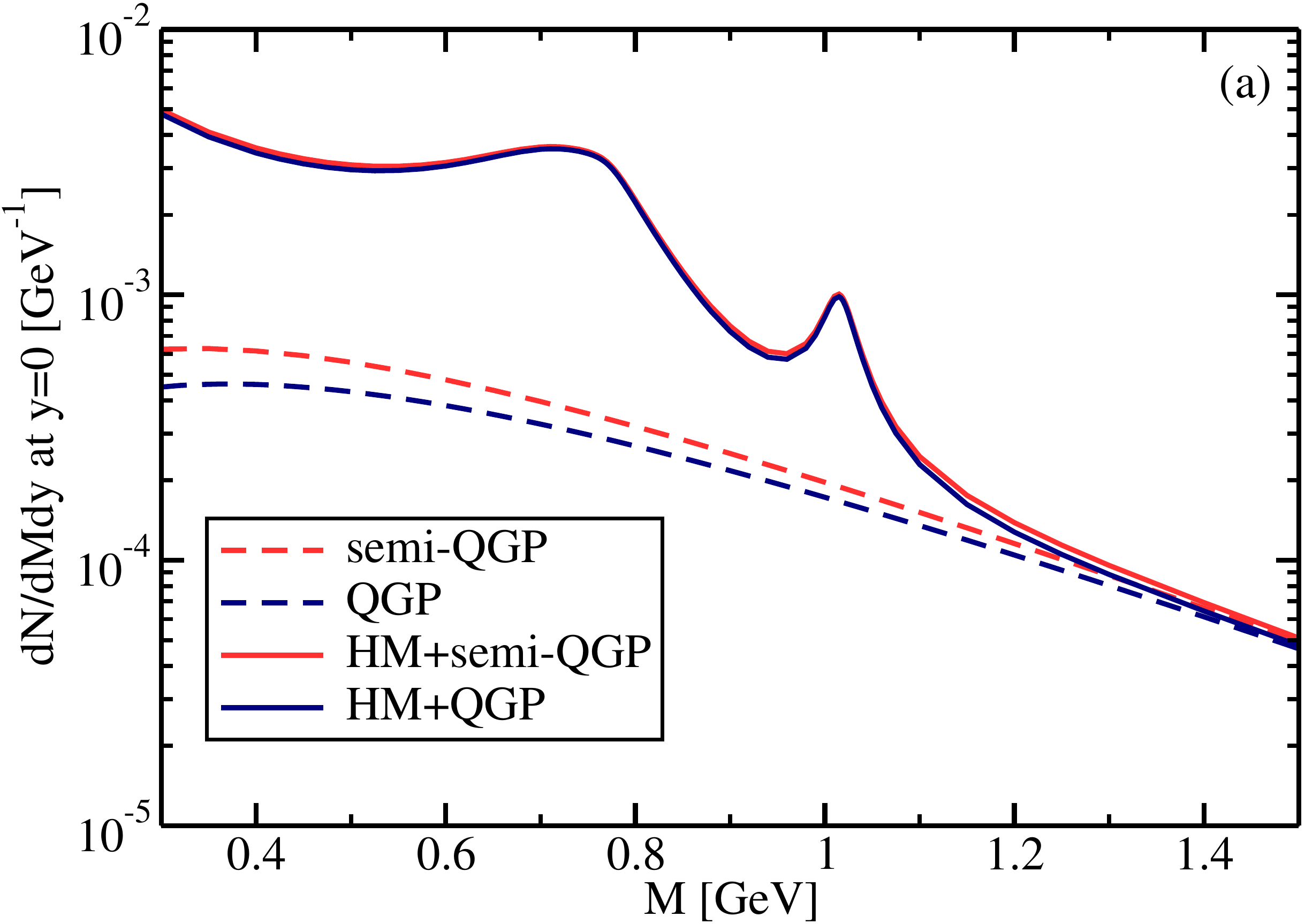} \\
\includegraphics[width=0.36\textwidth]{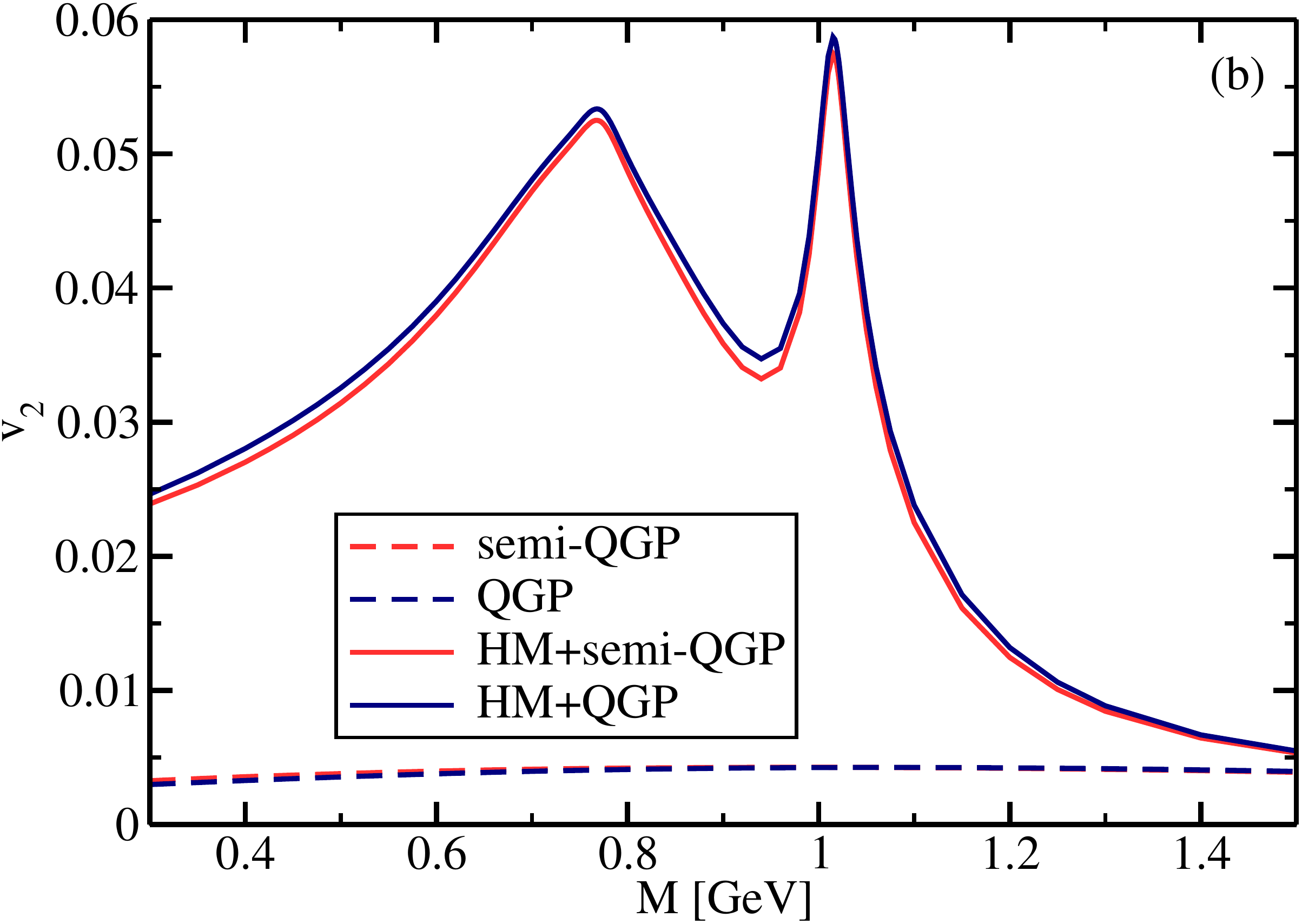}
\end{tabular}
\caption{ Dilepton yield (a) and elliptic flow (b)
computed using \textsc{music}, from
the semi-QGP and QGP, plus hadronic matter (HM). 
This calculation is for Au+Au collisions at the top RHIC energy, 
$\sqrt{s} = 200$GeV/A, in the 20-40\% centrality class.
}\label{fig:yield_v2_M}
\end{center}
\end{figure}
\begin{figure}[t]
\begin{center}
\vspace{0.4cm}
\begin{tabular} {c c}
\includegraphics[width=0.36\textwidth]{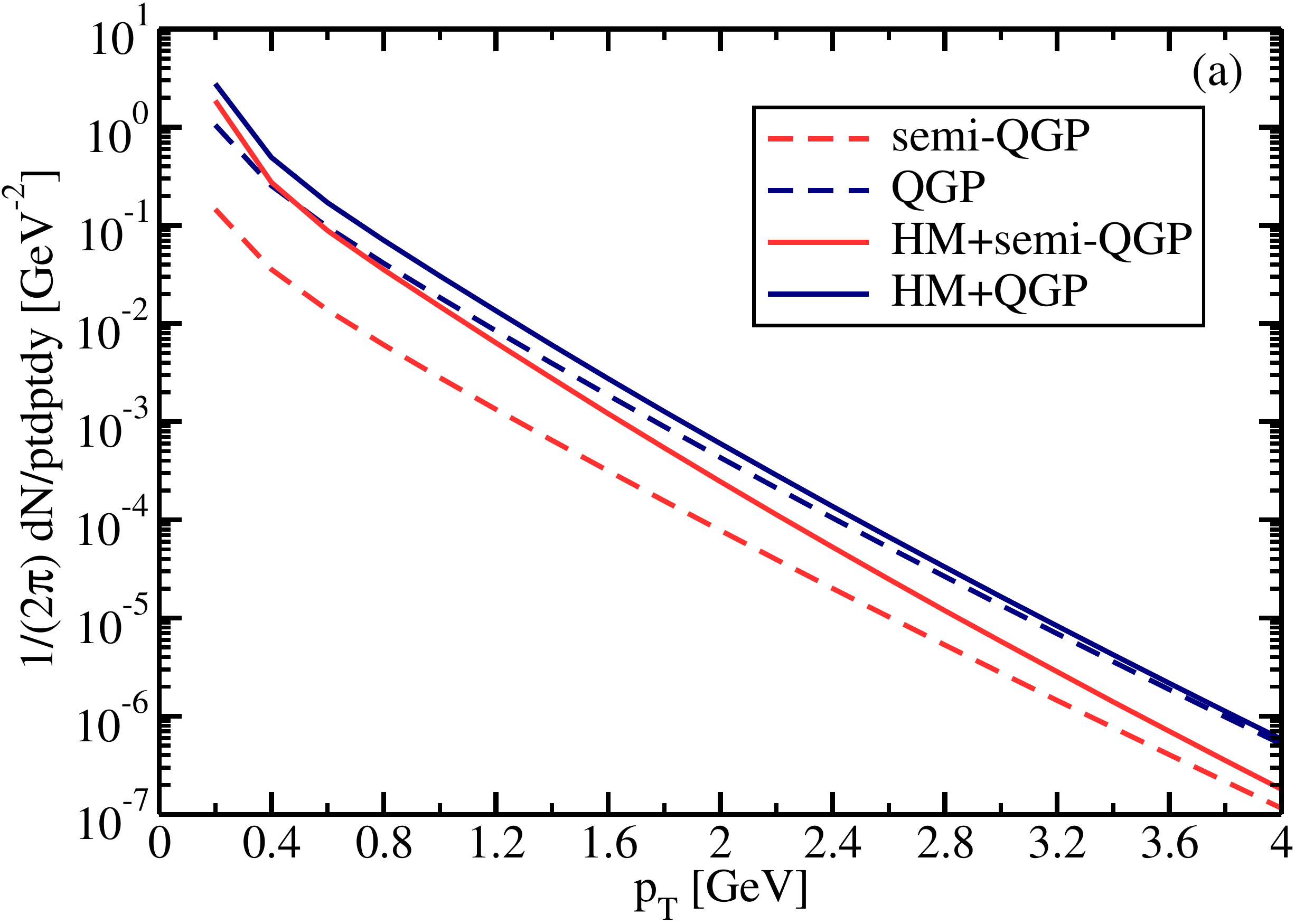}\\
\vspace{0.4cm}\\
\includegraphics[width=0.36\textwidth]{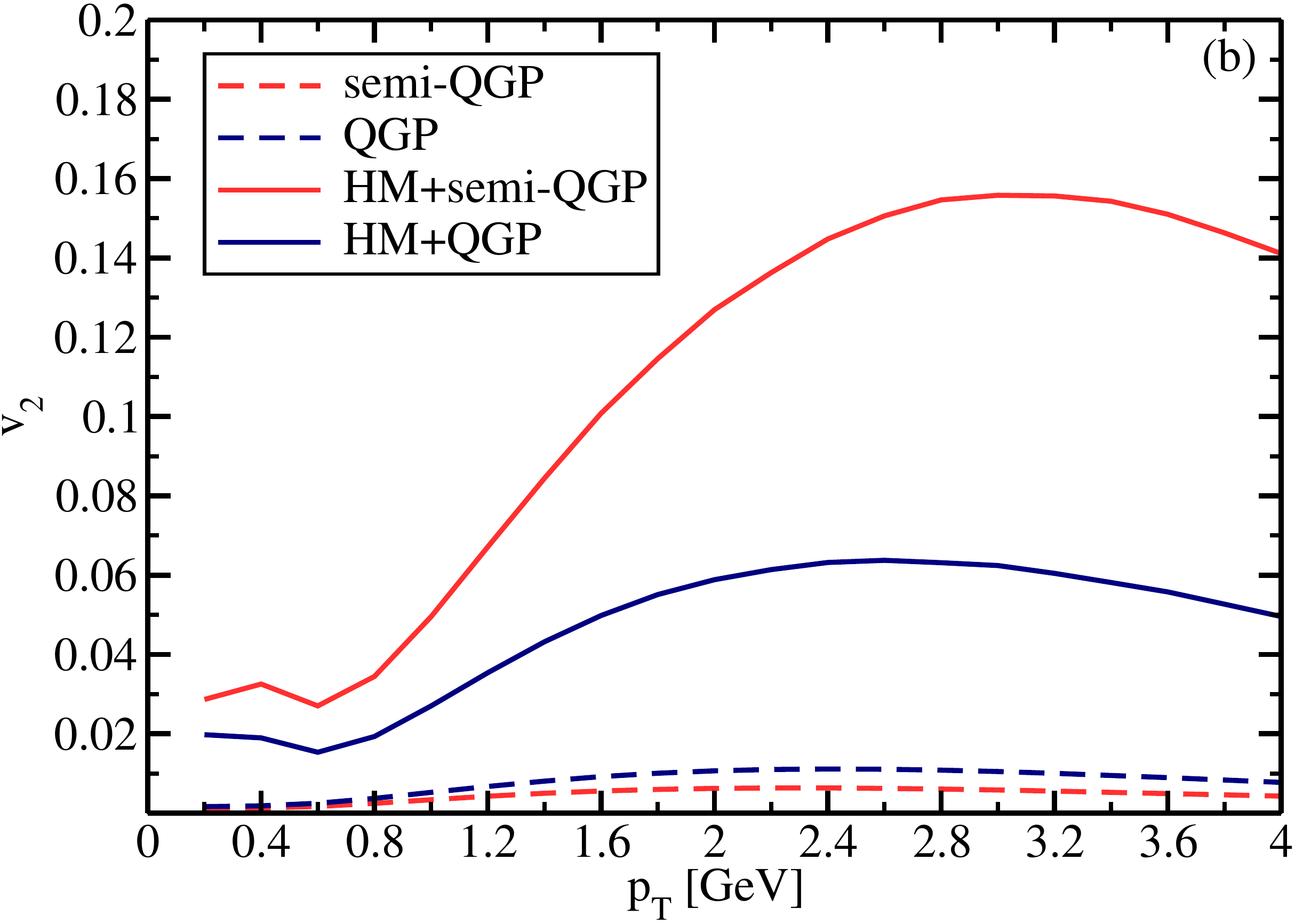}
\end{tabular}
\caption{ Photon yield (a) 
and elliptic flow (b) using \textsc{music},
from the semi-QGP and QGP, plus
hadronic matter (HM). As in Fig. (\ref{fig:yield_v2_pt}), 
this calculation is for Au+Au collisions 
at the top RHIC energy, in the 20-40\% centrality class.}
\label{fig:yield_v2_pt}
\end{center}
\end{figure}

Moving towards a qualitative estimate of the effects upon experiment,
we multiply the
full photon emission rate to $\sim e^2 g^2$
\cite{Aurenche:2000gf, Arnold:2002ja, *Arnold:2001ms, *Arnold:2002ja}
by the suppression factor which we find to leading logarithmic order,
$f_\gamma(Q)$ in Eq. (\ref{f_gamma}).
We use \textsc{music}, a 3+1 D hydrodynamic simulation 
\cite{Schenke:2010nt,Schenke:2010rr}. 
As the purpose of this study is to determine the global effect of 
rates in the semi-QGP, versus that in the usual QGP, 
we also include the hadronic rates for dileptons \cite{Vujanovic:2013jpa}
and photons \cite{Turbide:2003si}.  
We use ideal hydrodynamics for nucleus-nucleus collisions,
with $A = 200$ at RHIC energies, $\sqrt{s_{N N}} = 200$~GeV.

In ideal hydrodynamics, fluid dynamics is governed by the conservation equation
for the stress-energy tensor, 
$
\partial _{\mu }T^{\mu \nu }=0
$,
where
$
T^{\mu \nu }  =  (\varepsilon + P) u^{\mu }u^{\nu }-g^{\mu \nu}P
$;
$\varepsilon$ is the energy density, $P$ the thermodynamic
pressure and $u^{\mu}$ the fluid four-velocity.
The details regarding the numerical algorithm being used to solve the
hydrodynamic equations along with the initial and freeze-out conditions
are presented in Ref. \cite{Schenke:2010nt}. 

Fig. (\ref{fig:yield_v2_M}) shows the results for the dileptons.
There are slightly more dileptons from the semi-QGP than the usual QGP,
but below an invariant mass of $1.5$\,GeV, the total yield
is dominated by the hadronic matter. 
It might be possible to detect dileptons from the semi-QGP 
above $1.5$\,GeV.  The dilepton elliptic flow is small,
$v_2 \sim 0.01 - 0.06$, and is dominated by that from hadronic matter.

The results for photons, shown in Fig.~(\ref{fig:yield_v2_pt}),
are very different.  The suppression of color in the semi-QGP
greatly reduces the photon yield, Fig.~(\ref{fig:yield_v2_pt}a).
The $v_2$ of the semi-QGP is also reduced with respect to 
that of the QGP, Fig.~(\ref{fig:yield_v2_pt}b).

However, the {\it total} thermal photon $v_2$ is a 
yield-weighted average of the 
$v_2$ from the QGP and hadronic phases.
There is a competition between the change in the QGP yield 
and that of $v_2$: lowering the QGP $v_2$ lowers the 
thermal photon $v_2$, while a decrease in the yield from the QGP 
biases the thermal photon $v_2$ towards that from the hadronic phase, 
which is large. From Fig.~(\ref{fig:yield_v2_pt}b),
the latter wins, 
so that using semi-QGP rates significantly increases the 
total $v_2$ for thermal photons.

To make a detailed comparison to experiment, it is 
crucial to take into account
the contribution of prompt photons, produced through 
the collisions of hard partons.
Prompt photons can be computed using perturbative QCD, 
which work well in proton-proton collisions at next to leading order 
\cite{Adare:2012yt}. The dominant uncertainties of the perturbative 
calculation 
are the limited knowledge of the parton fragmentation functions into photons, 
and the dependence on the renormalization mass scale
\cite{Klasen:2013mga, *Klasen:2014xfa}.

There are further difficulties in extrapolating 
the contribution of prompt photons from proton-proton
to heavy ion collisions.  
Parton distribution functions are nuclear dependent, but
more importantly, 
parton fragmentation functions are medium-dependent
\cite{Arleo:2006xb, *Arleo:2008dn, *Arleo:2011gc}.
Photons with low
$p_T$ are produced predominantly by parton fragmentation.
In heavy ion collisions, experimentally it is found that
hadrons with $p_T$ up to $\approx 10$ GeV 
exhibit large elliptic flow.  
Thus a photon produced by the fragmentation of a hard jet should inherit
at least some of the elliptic flow of that jet.

Lastly, in heavy ion collisions at moderate $p_T$,
the number of photons produced by 
both perturbative and thermal mechanisms
appear to be significantly below that observed experimentally.
This may be due to an underestimate of rates in the hadronic medium
\cite{Steele:1996su, *Steele:1997tv, *Steele:1999hf, *Dusling:2006yv, *Dusling:2007su, *Dusling:2009ej, *Dusling:2009ej}
or to photons produced in the inital state, such as from the Color Glass
Condensate \cite{McLerran:2014hza}.

Previous computations in the semi-QGP show that like
photon production, processes involving color excitations,
such as the shear viscosity
\cite{Hidaka:2008dr, *Hidaka:2009hs, *Hidaka:2009xh, *Hidaka:2009ma}
and the collisional energy loss of heavy quarks \cite{Lin:2013efa},
are strongly suppressed near $T_c$.  
Thus we expect that the radiative energy loss of light quarks, which
involves the Landau-Pomeranchuk-Migdal effect,
is strongly suppressed near $T_c$.  Such computations are presently underway.

\begin{acknowledgements}

We thank H. van Hees, Y.-Q. Ma, L. McLerran, J. Qiu, R. Rapp, B. Schenke, 
W. Vogelsang, and I. Zahed for useful discussions.
C.G., S.J., J.-F.P. and G.V. are 
supported in part by the Natural Sciences 
and Engineering Research Council of Canada. 
Y.H.  is partially supported by JSPS KAKENHI Grants Numbers 24740184, 
and by  the RIKEN iTHES Project.
S.L. is supported by the RIKEN Foreign Postdoctoral Researchers Program.
J.-F.P. and G. V. acknowledge scholarships from 
Hydro- Quebec, FRQNT, and from the Canadian Institute of Nuclear Physics.
R.D.P.  is supported
by the U.S. Department of Energy under contract \#DE-AC02-98CH10886.
D. S. is supported by JSPS Strategic Young
Researcher Overseas Visits Program
for Accelerating Brain Circulation (No. R2411).
\end{acknowledgements}

\bibliography{photon}

\begin{thebibliography}{78}%
\makeatletter
\providecommand \@ifxundefined [1]{%
 \@ifx{#1\undefined}
}%
\providecommand \@ifnum [1]{%
 \ifnum #1\expandafter \@firstoftwo
 \else \expandafter \@secondoftwo
 \fi
}%
\providecommand \@ifx [1]{%
 \ifx #1\expandafter \@firstoftwo
 \else \expandafter \@secondoftwo
 \fi
}%
\providecommand \natexlab [1]{#1}%
\providecommand \enquote  [1]{``#1''}%
\providecommand \bibnamefont  [1]{#1}%
\providecommand \bibfnamefont [1]{#1}%
\providecommand \citenamefont [1]{#1}%
\providecommand \href@noop [0]{\@secondoftwo}%
\providecommand \href [0]{\begingroup \@sanitize@url \@href}%
\providecommand \@href[1]{\@@startlink{#1}\@@href}%
\providecommand \@@href[1]{\endgroup#1\@@endlink}%
\providecommand \@sanitize@url [0]{\catcode `\\12\catcode `\$12\catcode
  `\&12\catcode `\#12\catcode `\^12\catcode `\_12\catcode `\%12\relax}%
\providecommand \@@startlink[1]{}%
\providecommand \@@endlink[0]{}%
\providecommand \url  [0]{\begingroup\@sanitize@url \@url }%
\providecommand \@url [1]{\endgroup\@href {#1}{\urlprefix }}%
\providecommand \urlprefix  [0]{URL }%
\providecommand \Eprint [0]{\href }%
\providecommand \doibase [0]{http://dx.doi.org/}%
\providecommand \selectlanguage [0]{\@gobble}%
\providecommand \bibinfo  [0]{\@secondoftwo}%
\providecommand \bibfield  [0]{\@secondoftwo}%
\providecommand \translation [1]{[#1]}%
\providecommand \BibitemOpen [0]{}%
\providecommand \bibitemStop [0]{}%
\providecommand \bibitemNoStop [0]{.\EOS\space}%
\providecommand \EOS [0]{\spacefactor3000\relax}%
\providecommand \BibitemShut  [1]{\csname bibitem#1\endcsname}%
\let\auto@bib@innerbib\@empty
\bibitem [{\citenamefont {Haque}\ \emph {et~al.}(2014)\citenamefont {Haque},
  \citenamefont {Bandyopadhyay}, \citenamefont {Andersen}, \citenamefont
  {Mustafa}, \citenamefont {Strickland} \emph {et~al.}}]{Haque:2014rua}%
  \BibitemOpen
  \bibfield  {author} {\bibinfo {author} {\bibfnamefont {N.}~\bibnamefont
  {Haque}}, \bibinfo {author} {\bibfnamefont {A.}~\bibnamefont
  {Bandyopadhyay}}, \bibinfo {author} {\bibfnamefont {J.~O.}\ \bibnamefont
  {Andersen}}, \bibinfo {author} {\bibfnamefont {M.~G.}\ \bibnamefont
  {Mustafa}}, \bibinfo {author} {\bibfnamefont {M.}~\bibnamefont {Strickland}},
   \emph {et~al.},\ }\href@noop {} {\  (\bibinfo {year} {2014})},\ \Eprint
  {http://arxiv.org/abs/1402.6907} {arXiv:1402.6907 [hep-ph]} \BibitemShut
  {NoStop}%
\bibitem [{\citenamefont {Huovinen}\ and\ \citenamefont
  {Petreczky}(2010)}]{Huovinen:2009yb}%
  \BibitemOpen
  \bibfield  {author} {\bibinfo {author} {\bibfnamefont {P.}~\bibnamefont
  {Huovinen}}\ and\ \bibinfo {author} {\bibfnamefont {P.}~\bibnamefont
  {Petreczky}},\ }\href {\doibase 10.1016/j.nuclphysa.2010.02.015} {\bibfield
  {journal} {\bibinfo  {journal} {Nucl.Phys.}\ }\textbf {\bibinfo {volume}
  {A837}},\ \bibinfo {pages} {26} (\bibinfo {year} {2010})},\ \Eprint
  {http://arxiv.org/abs/0912.2541} {arXiv:0912.2541 [hep-ph]} \BibitemShut
  {NoStop}%
\bibitem [{\citenamefont {Andronic}\ \emph {et~al.}(2012)\citenamefont
  {Andronic}, \citenamefont {Braun-Munzinger}, \citenamefont {Stachel},\ and\
  \citenamefont {Winn}}]{Andronic:2012ut}%
  \BibitemOpen
  \bibfield  {author} {\bibinfo {author} {\bibfnamefont {A.}~\bibnamefont
  {Andronic}}, \bibinfo {author} {\bibfnamefont {P.}~\bibnamefont
  {Braun-Munzinger}}, \bibinfo {author} {\bibfnamefont {J.}~\bibnamefont
  {Stachel}}, \ and\ \bibinfo {author} {\bibfnamefont {M.}~\bibnamefont
  {Winn}},\ }\href {\doibase 10.1016/j.physletb.2012.10.001} {\bibfield
  {journal} {\bibinfo  {journal} {Phys.Lett.}\ }\textbf {\bibinfo {volume}
  {B718}},\ \bibinfo {pages} {80} (\bibinfo {year} {2012})},\ \Eprint
  {http://arxiv.org/abs/1201.0693} {arXiv:1201.0693 [nucl-th]} \BibitemShut
  {NoStop}%
\bibitem [{\citenamefont {Pisarski}(2006)}]{Pisarski:2006hz}%
  \BibitemOpen
  \bibfield  {author} {\bibinfo {author} {\bibfnamefont {R.~D.}\ \bibnamefont
  {Pisarski}},\ }\href {\doibase 10.1103/PhysRevD.74.121703} {\bibfield
  {journal} {\bibinfo  {journal} {Phys.Rev.}\ }\textbf {\bibinfo {volume}
  {D74}},\ \bibinfo {pages} {121703} (\bibinfo {year} {2006})},\ \Eprint
  {http://arxiv.org/abs/hep-ph/0608242} {arXiv:hep-ph/0608242 [hep-ph]}
  \BibitemShut {NoStop}%
\bibitem [{\citenamefont {Hidaka}\ and\ \citenamefont
  {Pisarski}(2008)}]{Hidaka:2008dr}%
  \BibitemOpen
  \bibfield  {author} {\bibinfo {author} {\bibfnamefont {Y.}~\bibnamefont
  {Hidaka}}\ and\ \bibinfo {author} {\bibfnamefont {R.~D.}\ \bibnamefont
  {Pisarski}},\ }\href {\doibase 10.1103/PhysRevD.78.071501} {\bibfield
  {journal} {\bibinfo  {journal} {Phys.Rev.}\ }\textbf {\bibinfo {volume}
  {D78}},\ \bibinfo {pages} {071501} (\bibinfo {year} {2008})},\ \Eprint
  {http://arxiv.org/abs/0803.0453} {arXiv:0803.0453 [hep-ph]} \BibitemShut
  {NoStop}%
\bibitem [{\citenamefont {Hidaka}\ and\ \citenamefont
  {Pisarski}(2009{\natexlab{a}})}]{Hidaka:2009hs}%
  \BibitemOpen
  \bibfield  {author} {\bibinfo {author} {\bibfnamefont {Y.}~\bibnamefont
  {Hidaka}}\ and\ \bibinfo {author} {\bibfnamefont {R.~D.}\ \bibnamefont
  {Pisarski}},\ }\href {\doibase 10.1103/PhysRevD.80.036004} {\bibfield
  {journal} {\bibinfo  {journal} {Phys.Rev.}\ }\textbf {\bibinfo {volume}
  {D80}},\ \bibinfo {pages} {036004} (\bibinfo {year} {2009}{\natexlab{a}})},\
  \Eprint {http://arxiv.org/abs/0906.1751} {arXiv:0906.1751 [hep-ph]}
  \BibitemShut {NoStop}%
\bibitem [{\citenamefont {Hidaka}\ and\ \citenamefont
  {Pisarski}(2009{\natexlab{b}})}]{Hidaka:2009xh}%
  \BibitemOpen
  \bibfield  {author} {\bibinfo {author} {\bibfnamefont {Y.}~\bibnamefont
  {Hidaka}}\ and\ \bibinfo {author} {\bibfnamefont {R.~D.}\ \bibnamefont
  {Pisarski}},\ }\href {\doibase 10.1103/PhysRevD.80.074504} {\bibfield
  {journal} {\bibinfo  {journal} {Phys.Rev.}\ }\textbf {\bibinfo {volume}
  {D80}},\ \bibinfo {pages} {074504} (\bibinfo {year} {2009}{\natexlab{b}})},\
  \Eprint {http://arxiv.org/abs/0907.4609} {arXiv:0907.4609 [hep-ph]}
  \BibitemShut {NoStop}%
\bibitem [{\citenamefont {Hidaka}\ and\ \citenamefont
  {Pisarski}(2010)}]{Hidaka:2009ma}%
  \BibitemOpen
  \bibfield  {author} {\bibinfo {author} {\bibfnamefont {Y.}~\bibnamefont
  {Hidaka}}\ and\ \bibinfo {author} {\bibfnamefont {R.~D.}\ \bibnamefont
  {Pisarski}},\ }\href {\doibase 10.1103/PhysRevD.81.076002} {\bibfield
  {journal} {\bibinfo  {journal} {Phys.Rev.}\ }\textbf {\bibinfo {volume}
  {D81}},\ \bibinfo {pages} {076002} (\bibinfo {year} {2010})},\ \Eprint
  {http://arxiv.org/abs/0912.0940} {arXiv:0912.0940 [hep-ph]} \BibitemShut
  {NoStop}%
\bibitem [{\citenamefont {Dumitru}\ \emph {et~al.}(2011)\citenamefont
  {Dumitru}, \citenamefont {Guo}, \citenamefont {Hidaka}, \citenamefont
  {Kortha\~ls Altes},\ and\ \citenamefont {Pisarski}}]{Dumitru:2010mj}%
  \BibitemOpen
  \bibfield  {author} {\bibinfo {author} {\bibfnamefont {A.}~\bibnamefont
  {Dumitru}}, \bibinfo {author} {\bibfnamefont {Y.}~\bibnamefont {Guo}},
  \bibinfo {author} {\bibfnamefont {Y.}~\bibnamefont {Hidaka}}, \bibinfo
  {author} {\bibfnamefont {C.~P.}\ \bibnamefont {Kortha\~ls Altes}}, \ and\
  \bibinfo {author} {\bibfnamefont {R.~D.}\ \bibnamefont {Pisarski}},\ }\href
  {\doibase 10.1103/PhysRevD.83.034022} {\bibfield  {journal} {\bibinfo
  {journal} {Phys.Rev.}\ }\textbf {\bibinfo {volume} {D83}},\ \bibinfo {pages}
  {034022} (\bibinfo {year} {2011})},\ \Eprint {http://arxiv.org/abs/1011.3820}
  {arXiv:1011.3820 [hep-ph]} \BibitemShut {NoStop}%
\bibitem [{\citenamefont {Dumitru}\ \emph {et~al.}(2012)\citenamefont
  {Dumitru}, \citenamefont {Guo}, \citenamefont {Hidaka}, \citenamefont
  {Altes},\ and\ \citenamefont {Pisarski}}]{Dumitru:2012fw}%
  \BibitemOpen
  \bibfield  {author} {\bibinfo {author} {\bibfnamefont {A.}~\bibnamefont
  {Dumitru}}, \bibinfo {author} {\bibfnamefont {Y.}~\bibnamefont {Guo}},
  \bibinfo {author} {\bibfnamefont {Y.}~\bibnamefont {Hidaka}}, \bibinfo
  {author} {\bibfnamefont {C.~P.~K.}\ \bibnamefont {Altes}}, \ and\ \bibinfo
  {author} {\bibfnamefont {R.~D.}\ \bibnamefont {Pisarski}},\ }\href {\doibase
  10.1103/PhysRevD.86.105017} {\bibfield  {journal} {\bibinfo  {journal}
  {Phys.Rev.}\ }\textbf {\bibinfo {volume} {D86}},\ \bibinfo {pages} {105017}
  (\bibinfo {year} {2012})},\ \Eprint {http://arxiv.org/abs/1205.0137}
  {arXiv:1205.0137 [hep-ph]} \BibitemShut {NoStop}%
\bibitem [{\citenamefont {Kashiwa}\ \emph {et~al.}(2012)\citenamefont
  {Kashiwa}, \citenamefont {Pisarski},\ and\ \citenamefont
  {Skokov}}]{Kashiwa:2012wa}%
  \BibitemOpen
  \bibfield  {author} {\bibinfo {author} {\bibfnamefont {K.}~\bibnamefont
  {Kashiwa}}, \bibinfo {author} {\bibfnamefont {R.~D.}\ \bibnamefont
  {Pisarski}}, \ and\ \bibinfo {author} {\bibfnamefont {V.~V.}\ \bibnamefont
  {Skokov}},\ }\href {\doibase 10.1103/PhysRevD.85.114029} {\bibfield
  {journal} {\bibinfo  {journal} {Phys.Rev.}\ }\textbf {\bibinfo {volume}
  {D85}},\ \bibinfo {pages} {114029} (\bibinfo {year} {2012})},\ \Eprint
  {http://arxiv.org/abs/1205.0545} {arXiv:1205.0545 [hep-ph]} \BibitemShut
  {NoStop}%
\bibitem [{\citenamefont {Pisarski}\ and\ \citenamefont
  {Skokov}(2012)}]{Pisarski:2012bj}%
  \BibitemOpen
  \bibfield  {author} {\bibinfo {author} {\bibfnamefont {R.~D.}\ \bibnamefont
  {Pisarski}}\ and\ \bibinfo {author} {\bibfnamefont {V.~V.}\ \bibnamefont
  {Skokov}},\ }\href {\doibase 10.1103/PhysRevD.86.081701} {\bibfield
  {journal} {\bibinfo  {journal} {Phys.Rev.}\ }\textbf {\bibinfo {volume}
  {D86}},\ \bibinfo {pages} {081701} (\bibinfo {year} {2012})},\ \Eprint
  {http://arxiv.org/abs/1206.1329} {arXiv:1206.1329 [hep-th]} \BibitemShut
  {NoStop}%
\bibitem [{\citenamefont {Kashiwa}\ and\ \citenamefont
  {Pisarski}(2013)}]{Kashiwa:2013rm}%
  \BibitemOpen
  \bibfield  {author} {\bibinfo {author} {\bibfnamefont {K.}~\bibnamefont
  {Kashiwa}}\ and\ \bibinfo {author} {\bibfnamefont {R.~D.}\ \bibnamefont
  {Pisarski}},\ }\href {\doibase 10.1103/PhysRevD.87.096009} {\bibfield
  {journal} {\bibinfo  {journal} {Phys.Rev.}\ }\textbf {\bibinfo {volume}
  {D87}},\ \bibinfo {pages} {096009} (\bibinfo {year} {2013})},\ \Eprint
  {http://arxiv.org/abs/1301.5344} {arXiv:1301.5344 [hep-ph]} \BibitemShut
  {NoStop}%
\bibitem [{\citenamefont {Lin}\ \emph {et~al.}(2013{\natexlab{a}})\citenamefont
  {Lin}, \citenamefont {Pisarski},\ and\ \citenamefont {Skokov}}]{Lin:2013qu}%
  \BibitemOpen
  \bibfield  {author} {\bibinfo {author} {\bibfnamefont {S.}~\bibnamefont
  {Lin}}, \bibinfo {author} {\bibfnamefont {R.~D.}\ \bibnamefont {Pisarski}}, \
  and\ \bibinfo {author} {\bibfnamefont {V.~V.}\ \bibnamefont {Skokov}},\
  }\href {\doibase 10.1103/PhysRevD.87.105002} {\bibfield  {journal} {\bibinfo
  {journal} {Phys.Rev.}\ }\textbf {\bibinfo {volume} {D87}},\ \bibinfo {pages}
  {105002} (\bibinfo {year} {2013}{\natexlab{a}})},\ \Eprint
  {http://arxiv.org/abs/1301.7432} {arXiv:1301.7432 [hep-ph]} \BibitemShut
  {NoStop}%
\bibitem [{\citenamefont {Lin}\ \emph {et~al.}(2013{\natexlab{b}})\citenamefont
  {Lin}, \citenamefont {Pisarski},\ and\ \citenamefont {Skokov}}]{Lin:2013efa}%
  \BibitemOpen
  \bibfield  {author} {\bibinfo {author} {\bibfnamefont {S.}~\bibnamefont
  {Lin}}, \bibinfo {author} {\bibfnamefont {R.~D.}\ \bibnamefont {Pisarski}}, \
  and\ \bibinfo {author} {\bibfnamefont {V.~V.}\ \bibnamefont {Skokov}},\
  }\href {\doibase 10.1016/j.physletb.2014.01.043} {\  (\bibinfo {year}
  {2013}{\natexlab{b}}),\ 10.1016/j.physletb.2014.01.043},\ \Eprint
  {http://arxiv.org/abs/1312.3340} {arXiv:1312.3340 [hep-ph]} \BibitemShut
  {NoStop}%
\bibitem [{\citenamefont {Bazavov}\ \emph {et~al.}(2009)\citenamefont
  {Bazavov}, \citenamefont {Bhattacharya}, \citenamefont {Cheng}, \citenamefont
  {Christ}, \citenamefont {DeTar} \emph {et~al.}}]{Bazavov:2009zn}%
  \BibitemOpen
  \bibfield  {author} {\bibinfo {author} {\bibfnamefont {A.}~\bibnamefont
  {Bazavov}}, \bibinfo {author} {\bibfnamefont {T.}~\bibnamefont
  {Bhattacharya}}, \bibinfo {author} {\bibfnamefont {M.}~\bibnamefont {Cheng}},
  \bibinfo {author} {\bibfnamefont {N.}~\bibnamefont {Christ}}, \bibinfo
  {author} {\bibfnamefont {C.}~\bibnamefont {DeTar}},  \emph {et~al.},\ }\href
  {\doibase 10.1103/PhysRevD.80.014504} {\bibfield  {journal} {\bibinfo
  {journal} {Phys.Rev.}\ }\textbf {\bibinfo {volume} {D80}},\ \bibinfo {pages}
  {014504} (\bibinfo {year} {2009})},\ \Eprint {http://arxiv.org/abs/0903.4379}
  {arXiv:0903.4379 [hep-lat]} \BibitemShut {NoStop}%
\bibitem [{\citenamefont {DeTar}\ and\ \citenamefont
  {Heller}(2009)}]{DeTar:2009ef}%
  \BibitemOpen
  \bibfield  {author} {\bibinfo {author} {\bibfnamefont {C.}~\bibnamefont
  {DeTar}}\ and\ \bibinfo {author} {\bibfnamefont {U.}~\bibnamefont {Heller}},\
  }\href {\doibase 10.1140/epja/i2009-10825-3} {\bibfield  {journal} {\bibinfo
  {journal} {Eur.Phys.J.}\ }\textbf {\bibinfo {volume} {A41}},\ \bibinfo
  {pages} {405} (\bibinfo {year} {2009})},\ \Eprint
  {http://arxiv.org/abs/0905.2949} {arXiv:0905.2949 [hep-lat]} \BibitemShut
  {NoStop}%
\bibitem [{\citenamefont {Fodor}\ and\ \citenamefont
  {Katz}(2009)}]{Fodor:2009ax}%
  \BibitemOpen
  \bibfield  {author} {\bibinfo {author} {\bibfnamefont {Z.}~\bibnamefont
  {Fodor}}\ and\ \bibinfo {author} {\bibfnamefont {S.}~\bibnamefont {Katz}},\
  }\href@noop {} {\  (\bibinfo {year} {2009})},\ \Eprint
  {http://arxiv.org/abs/0908.3341} {arXiv:0908.3341 [hep-ph]} \BibitemShut
  {NoStop}%
\bibitem [{\citenamefont {Petreczky}(2012)}]{Petreczky:2012rq}%
  \BibitemOpen
  \bibfield  {author} {\bibinfo {author} {\bibfnamefont {P.}~\bibnamefont
  {Petreczky}},\ }\href {\doibase 10.1088/0954-3899/39/9/093002} {\bibfield
  {journal} {\bibinfo  {journal} {J.Phys.}\ }\textbf {\bibinfo {volume}
  {G39}},\ \bibinfo {pages} {093002} (\bibinfo {year} {2012})},\ \Eprint
  {http://arxiv.org/abs/1203.5320} {arXiv:1203.5320 [hep-lat]} \BibitemShut
  {NoStop}%
\bibitem [{\citenamefont {Borsanyi}\ \emph {et~al.}(2014)\citenamefont
  {Borsanyi}, \citenamefont {Fodor}, \citenamefont {Hoelbling}, \citenamefont
  {Katz}, \citenamefont {Krieg} \emph {et~al.}}]{Borsanyi:2013bia}%
  \BibitemOpen
  \bibfield  {author} {\bibinfo {author} {\bibfnamefont {S.}~\bibnamefont
  {Borsanyi}}, \bibinfo {author} {\bibfnamefont {Z.}~\bibnamefont {Fodor}},
  \bibinfo {author} {\bibfnamefont {C.}~\bibnamefont {Hoelbling}}, \bibinfo
  {author} {\bibfnamefont {S.~D.}\ \bibnamefont {Katz}}, \bibinfo {author}
  {\bibfnamefont {S.}~\bibnamefont {Krieg}},  \emph {et~al.},\ }\href {\doibase
  10.1016/j.physletb.2014.01.007} {\bibfield  {journal} {\bibinfo  {journal}
  {Phys.Lett.}\ }\textbf {\bibinfo {volume} {B370}},\ \bibinfo {pages} {99}
  (\bibinfo {year} {2014})},\ \Eprint {http://arxiv.org/abs/1309.5258}
  {arXiv:1309.5258 [hep-lat]} \BibitemShut {NoStop}%
\bibitem [{\citenamefont {Bhattacharya}\ \emph {et~al.}(2014)\citenamefont
  {Bhattacharya}, \citenamefont {Buchoff}, \citenamefont {Christ},
  \citenamefont {Ding}, \citenamefont {Gupta} \emph
  {et~al.}}]{Bhattacharya:2014ara}%
  \BibitemOpen
  \bibfield  {author} {\bibinfo {author} {\bibfnamefont {T.}~\bibnamefont
  {Bhattacharya}}, \bibinfo {author} {\bibfnamefont {M.~I.}\ \bibnamefont
  {Buchoff}}, \bibinfo {author} {\bibfnamefont {N.~H.}\ \bibnamefont {Christ}},
  \bibinfo {author} {\bibfnamefont {H.~T.}\ \bibnamefont {Ding}}, \bibinfo
  {author} {\bibfnamefont {R.}~\bibnamefont {Gupta}},  \emph {et~al.},\
  }\href@noop {} {\  (\bibinfo {year} {2014})},\ \Eprint
  {http://arxiv.org/abs/1402.5175} {arXiv:1402.5175 [hep-lat]} \BibitemShut
  {NoStop}%
\bibitem [{\citenamefont {Sharma}(2013)}]{Sharma:2013hsa}%
  \BibitemOpen
  \bibfield  {author} {\bibinfo {author} {\bibfnamefont {S.}~\bibnamefont
  {Sharma}},\ }\href {\doibase 10.1155/2013/452978} {\bibfield  {journal}
  {\bibinfo  {journal} {Adv.High Energy Phys.}\ }\textbf {\bibinfo {volume}
  {2013}},\ \bibinfo {pages} {452978} (\bibinfo {year} {2013})},\ \Eprint
  {http://arxiv.org/abs/1403.2102} {arXiv:1403.2102 [hep-lat]} \BibitemShut
  {NoStop}%
\bibitem [{\citenamefont {Heinz}(2009)}]{Heinz:2009xj}%
  \BibitemOpen
  \bibfield  {author} {\bibinfo {author} {\bibfnamefont {U.~W.}\ \bibnamefont
  {Heinz}},\ }\href@noop {} {\  (\bibinfo {year} {2009})},\ \Eprint
  {http://arxiv.org/abs/0901.4355} {arXiv:0901.4355 [nucl-th]} \BibitemShut
  {NoStop}%
\bibitem [{\citenamefont {Heinz}\ and\ \citenamefont
  {Snellings}(2013)}]{Heinz:2013th}%
  \BibitemOpen
  \bibfield  {author} {\bibinfo {author} {\bibfnamefont {U.}~\bibnamefont
  {Heinz}}\ and\ \bibinfo {author} {\bibfnamefont {R.}~\bibnamefont
  {Snellings}},\ }\href {\doibase 10.1146/annurev-nucl-102212-170540}
  {\bibfield  {journal} {\bibinfo  {journal} {Ann.Rev.Nucl.Part.Sci.}\ }\textbf
  {\bibinfo {volume} {63}},\ \bibinfo {pages} {123} (\bibinfo {year} {2013})},\
  \Eprint {http://arxiv.org/abs/1301.2826} {arXiv:1301.2826 [nucl-th]}
  \BibitemShut {NoStop}%
\bibitem [{\citenamefont {Gale}\ \emph {et~al.}(2013)\citenamefont {Gale},
  \citenamefont {Jeon},\ and\ \citenamefont {Schenke}}]{Gale:2013da}%
  \BibitemOpen
  \bibfield  {author} {\bibinfo {author} {\bibfnamefont {C.}~\bibnamefont
  {Gale}}, \bibinfo {author} {\bibfnamefont {S.}~\bibnamefont {Jeon}}, \ and\
  \bibinfo {author} {\bibfnamefont {B.}~\bibnamefont {Schenke}},\ }\href
  {\doibase 10.1142/S0217751X13400113} {\bibfield  {journal} {\bibinfo
  {journal} {Int.J.Mod.Phys.}\ }\textbf {\bibinfo {volume} {A28}},\ \bibinfo
  {pages} {1340011} (\bibinfo {year} {2013})},\ \Eprint
  {http://arxiv.org/abs/1301.5893} {arXiv:1301.5893 [nucl-th]} \BibitemShut
  {NoStop}%
\bibitem [{\citenamefont {Schenke}\ \emph {et~al.}(2010)\citenamefont
  {Schenke}, \citenamefont {Jeon},\ and\ \citenamefont
  {Gale}}]{Schenke:2010nt}%
  \BibitemOpen
  \bibfield  {author} {\bibinfo {author} {\bibfnamefont {B.}~\bibnamefont
  {Schenke}}, \bibinfo {author} {\bibfnamefont {S.}~\bibnamefont {Jeon}}, \
  and\ \bibinfo {author} {\bibfnamefont {C.}~\bibnamefont {Gale}},\ }\href
  {\doibase 10.1103/PhysRevC.82.014903} {\bibfield  {journal} {\bibinfo
  {journal} {Phys.Rev.}\ }\textbf {\bibinfo {volume} {C82}},\ \bibinfo {pages}
  {014903} (\bibinfo {year} {2010})},\ \Eprint {http://arxiv.org/abs/1004.1408}
  {arXiv:1004.1408 [hep-ph]} \BibitemShut {NoStop}%
\bibitem [{\citenamefont {Schenke}\ \emph {et~al.}(2011)\citenamefont
  {Schenke}, \citenamefont {Jeon},\ and\ \citenamefont
  {Gale}}]{Schenke:2010rr}%
  \BibitemOpen
  \bibfield  {author} {\bibinfo {author} {\bibfnamefont {B.}~\bibnamefont
  {Schenke}}, \bibinfo {author} {\bibfnamefont {S.}~\bibnamefont {Jeon}}, \
  and\ \bibinfo {author} {\bibfnamefont {C.}~\bibnamefont {Gale}},\ }\href
  {\doibase 10.1103/PhysRevLett.106.042301} {\bibfield  {journal} {\bibinfo
  {journal} {Phys.Rev.Lett.}\ }\textbf {\bibinfo {volume} {106}},\ \bibinfo
  {pages} {042301} (\bibinfo {year} {2011})},\ \Eprint
  {http://arxiv.org/abs/1009.3244} {arXiv:1009.3244 [hep-ph]} \BibitemShut
  {NoStop}%
\bibitem [{\citenamefont {Baier}\ \emph {et~al.}(1992)\citenamefont {Baier},
  \citenamefont {Nakkagawa}, \citenamefont {Niegawa},\ and\ \citenamefont
  {Redlich}}]{Baier:1991em}%
  \BibitemOpen
  \bibfield  {author} {\bibinfo {author} {\bibfnamefont {R.}~\bibnamefont
  {Baier}}, \bibinfo {author} {\bibfnamefont {H.}~\bibnamefont {Nakkagawa}},
  \bibinfo {author} {\bibfnamefont {A.}~\bibnamefont {Niegawa}}, \ and\
  \bibinfo {author} {\bibfnamefont {K.}~\bibnamefont {Redlich}},\ }\href
  {\doibase 10.1007/BF01625902} {\bibfield  {journal} {\bibinfo  {journal}
  {Z.Phys.}\ }\textbf {\bibinfo {volume} {C53}},\ \bibinfo {pages} {433}
  (\bibinfo {year} {1992})}\BibitemShut {NoStop}%
\bibitem [{\citenamefont {Kapusta}\ \emph {et~al.}(1991)\citenamefont
  {Kapusta}, \citenamefont {Lichard},\ and\ \citenamefont
  {Seibert}}]{Kapusta:1991qp}%
  \BibitemOpen
  \bibfield  {author} {\bibinfo {author} {\bibfnamefont {J.~I.}\ \bibnamefont
  {Kapusta}}, \bibinfo {author} {\bibfnamefont {P.}~\bibnamefont {Lichard}}, \
  and\ \bibinfo {author} {\bibfnamefont {D.}~\bibnamefont {Seibert}},\ }\href
  {\doibase 10.1103/PhysRevD.47.4171, 10.1103/PhysRevD.44.2774} {\bibfield
  {journal} {\bibinfo  {journal} {Phys.Rev.}\ }\textbf {\bibinfo {volume}
  {D44}},\ \bibinfo {pages} {2774} (\bibinfo {year} {1991})}\BibitemShut
  {NoStop}%
\bibitem [{\citenamefont {Steele}\ \emph {et~al.}(1996)\citenamefont {Steele},
  \citenamefont {Yamagishi},\ and\ \citenamefont {Zahed}}]{Steele:1996su}%
  \BibitemOpen
  \bibfield  {author} {\bibinfo {author} {\bibfnamefont {J.~V.}\ \bibnamefont
  {Steele}}, \bibinfo {author} {\bibfnamefont {H.}~\bibnamefont {Yamagishi}}, \
  and\ \bibinfo {author} {\bibfnamefont {I.}~\bibnamefont {Zahed}},\ }\href
  {\doibase 10.1016/0370-2693(96)00802-7} {\bibfield  {journal} {\bibinfo
  {journal} {Phys.Lett.}\ }\textbf {\bibinfo {volume} {B384}},\ \bibinfo
  {pages} {255} (\bibinfo {year} {1996})},\ \Eprint
  {http://arxiv.org/abs/hep-ph/9603290} {arXiv:hep-ph/9603290 [hep-ph]}
  \BibitemShut {NoStop}%
\bibitem [{\citenamefont {Steele}\ \emph {et~al.}(1997)\citenamefont {Steele},
  \citenamefont {Yamagishi},\ and\ \citenamefont {Zahed}}]{Steele:1997tv}%
  \BibitemOpen
  \bibfield  {author} {\bibinfo {author} {\bibfnamefont {J.~V.}\ \bibnamefont
  {Steele}}, \bibinfo {author} {\bibfnamefont {H.}~\bibnamefont {Yamagishi}}, \
  and\ \bibinfo {author} {\bibfnamefont {I.}~\bibnamefont {Zahed}},\ }\href
  {\doibase 10.1103/PhysRevD.56.5605} {\bibfield  {journal} {\bibinfo
  {journal} {Phys.Rev.}\ }\textbf {\bibinfo {volume} {D56}},\ \bibinfo {pages}
  {5605} (\bibinfo {year} {1997})},\ \Eprint
  {http://arxiv.org/abs/hep-ph/9704414} {arXiv:hep-ph/9704414 [hep-ph]}
  \BibitemShut {NoStop}%
\bibitem [{\citenamefont {Steele}\ and\ \citenamefont
  {Zahed}(1999)}]{Steele:1999hf}%
  \BibitemOpen
  \bibfield  {author} {\bibinfo {author} {\bibfnamefont {J.~V.}\ \bibnamefont
  {Steele}}\ and\ \bibinfo {author} {\bibfnamefont {I.}~\bibnamefont {Zahed}},\
  }\href {\doibase 10.1103/PhysRevD.60.037502} {\bibfield  {journal} {\bibinfo
  {journal} {Phys.Rev.}\ }\textbf {\bibinfo {volume} {D60}},\ \bibinfo {pages}
  {037502} (\bibinfo {year} {1999})},\ \Eprint
  {http://arxiv.org/abs/hep-ph/9901385} {arXiv:hep-ph/9901385 [hep-ph]}
  \BibitemShut {NoStop}%
\bibitem [{\citenamefont {Dusling}\ \emph {et~al.}(2007)\citenamefont
  {Dusling}, \citenamefont {Teaney},\ and\ \citenamefont
  {Zahed}}]{Dusling:2006yv}%
  \BibitemOpen
  \bibfield  {author} {\bibinfo {author} {\bibfnamefont {K.}~\bibnamefont
  {Dusling}}, \bibinfo {author} {\bibfnamefont {D.}~\bibnamefont {Teaney}}, \
  and\ \bibinfo {author} {\bibfnamefont {I.}~\bibnamefont {Zahed}},\ }\href
  {\doibase 10.1103/PhysRevC.75.024908} {\bibfield  {journal} {\bibinfo
  {journal} {Phys.Rev.}\ }\textbf {\bibinfo {volume} {C75}},\ \bibinfo {pages}
  {024908} (\bibinfo {year} {2007})},\ \Eprint
  {http://arxiv.org/abs/nucl-th/0604071} {arXiv:nucl-th/0604071 [nucl-th]}
  \BibitemShut {NoStop}%
\bibitem [{\citenamefont {Dusling}\ and\ \citenamefont
  {Zahed}(2009)}]{Dusling:2007su}%
  \BibitemOpen
  \bibfield  {author} {\bibinfo {author} {\bibfnamefont {K.}~\bibnamefont
  {Dusling}}\ and\ \bibinfo {author} {\bibfnamefont {I.}~\bibnamefont
  {Zahed}},\ }\href {\doibase 10.1016/j.nuclphysa.2009.04.013} {\bibfield
  {journal} {\bibinfo  {journal} {Nucl.Phys.}\ }\textbf {\bibinfo {volume}
  {A825}},\ \bibinfo {pages} {212} (\bibinfo {year} {2009})},\ \Eprint
  {http://arxiv.org/abs/0712.1982} {arXiv:0712.1982 [nucl-th]} \BibitemShut
  {NoStop}%
\bibitem [{\citenamefont {Dusling}\ and\ \citenamefont
  {Zahed}(2010)}]{Dusling:2009ej}%
  \BibitemOpen
  \bibfield  {author} {\bibinfo {author} {\bibfnamefont {K.}~\bibnamefont
  {Dusling}}\ and\ \bibinfo {author} {\bibfnamefont {I.}~\bibnamefont
  {Zahed}},\ }\href {\doibase 10.1103/PhysRevC.82.054909} {\bibfield  {journal}
  {\bibinfo  {journal} {Phys.Rev.}\ }\textbf {\bibinfo {volume} {C82}},\
  \bibinfo {pages} {054909} (\bibinfo {year} {2010})},\ \Eprint
  {http://arxiv.org/abs/0911.2426} {arXiv:0911.2426 [nucl-th]} \BibitemShut
  {NoStop}%
\bibitem [{\citenamefont {Lee}\ \emph {et~al.}(1999)\citenamefont {Lee},
  \citenamefont {Wirstam}, \citenamefont {Zahed},\ and\ \citenamefont
  {Hansson}}]{Lee:1998nz}%
  \BibitemOpen
  \bibfield  {author} {\bibinfo {author} {\bibfnamefont {C.}~\bibnamefont
  {Lee}}, \bibinfo {author} {\bibfnamefont {J.}~\bibnamefont {Wirstam}},
  \bibinfo {author} {\bibfnamefont {I.}~\bibnamefont {Zahed}}, \ and\ \bibinfo
  {author} {\bibfnamefont {T.}~\bibnamefont {Hansson}},\ }\href {\doibase
  10.1016/S0370-2693(99)00061-1} {\bibfield  {journal} {\bibinfo  {journal}
  {Phys.Lett.}\ }\textbf {\bibinfo {volume} {B448}},\ \bibinfo {pages} {168}
  (\bibinfo {year} {1999})},\ \Eprint {http://arxiv.org/abs/hep-ph/9809440}
  {arXiv:hep-ph/9809440 [hep-ph]} \BibitemShut {NoStop}%
\bibitem [{\citenamefont {Aurenche}\ \emph {et~al.}(2000)\citenamefont
  {Aurenche}, \citenamefont {Gelis},\ and\ \citenamefont
  {Zaraket}}]{Aurenche:2000gf}%
  \BibitemOpen
  \bibfield  {author} {\bibinfo {author} {\bibfnamefont {P.}~\bibnamefont
  {Aurenche}}, \bibinfo {author} {\bibfnamefont {F.}~\bibnamefont {Gelis}}, \
  and\ \bibinfo {author} {\bibfnamefont {H.}~\bibnamefont {Zaraket}},\ }\href
  {\doibase 10.1103/PhysRevD.62.096012} {\bibfield  {journal} {\bibinfo
  {journal} {Phys.Rev.}\ }\textbf {\bibinfo {volume} {D62}},\ \bibinfo {pages}
  {096012} (\bibinfo {year} {2000})},\ \Eprint
  {http://arxiv.org/abs/hep-ph/0003326} {arXiv:hep-ph/0003326 [hep-ph]}
  \BibitemShut {NoStop}%
\bibitem [{\citenamefont {Arnold}\ \emph {et~al.}(2002)\citenamefont {Arnold},
  \citenamefont {Moore},\ and\ \citenamefont {Yaffe}}]{Arnold:2002ja}%
  \BibitemOpen
  \bibfield  {author} {\bibinfo {author} {\bibfnamefont {P.~B.}\ \bibnamefont
  {Arnold}}, \bibinfo {author} {\bibfnamefont {G.~D.}\ \bibnamefont {Moore}}, \
  and\ \bibinfo {author} {\bibfnamefont {L.~G.}\ \bibnamefont {Yaffe}},\ }\href
  {\doibase 10.1088/1126-6708/2002/06/030} {\bibfield  {journal} {\bibinfo
  {journal} {JHEP}\ }\textbf {\bibinfo {volume} {0206}},\ \bibinfo {pages}
  {030} (\bibinfo {year} {2002})},\ \Eprint
  {http://arxiv.org/abs/hep-ph/0204343} {arXiv:hep-ph/0204343 [hep-ph]}
  \BibitemShut {NoStop}%
\bibitem [{\citenamefont {Arnold}\ \emph {et~al.}(2001)\citenamefont {Arnold},
  \citenamefont {Moore},\ and\ \citenamefont {Yaffe}}]{Arnold:2001ms}%
  \BibitemOpen
  \bibfield  {author} {\bibinfo {author} {\bibfnamefont {P.~B.}\ \bibnamefont
  {Arnold}}, \bibinfo {author} {\bibfnamefont {G.~D.}\ \bibnamefont {Moore}}, \
  and\ \bibinfo {author} {\bibfnamefont {L.~G.}\ \bibnamefont {Yaffe}},\ }\href
  {\doibase 10.1088/1126-6708/2001/12/009} {\bibfield  {journal} {\bibinfo
  {journal} {JHEP}\ }\textbf {\bibinfo {volume} {0112}},\ \bibinfo {pages}
  {009} (\bibinfo {year} {2001})},\ \Eprint
  {http://arxiv.org/abs/hep-ph/0111107} {arXiv:hep-ph/0111107 [hep-ph]}
  \BibitemShut {NoStop}%
\bibitem [{\citenamefont {Turbide}\ \emph {et~al.}(2004)\citenamefont
  {Turbide}, \citenamefont {Rapp},\ and\ \citenamefont
  {Gale}}]{Turbide:2003si}%
  \BibitemOpen
  \bibfield  {author} {\bibinfo {author} {\bibfnamefont {S.}~\bibnamefont
  {Turbide}}, \bibinfo {author} {\bibfnamefont {R.}~\bibnamefont {Rapp}}, \
  and\ \bibinfo {author} {\bibfnamefont {C.}~\bibnamefont {Gale}},\ }\href
  {\doibase 10.1103/PhysRevC.69.014903} {\bibfield  {journal} {\bibinfo
  {journal} {Phys.Rev.}\ }\textbf {\bibinfo {volume} {C69}},\ \bibinfo {pages}
  {014903} (\bibinfo {year} {2004})},\ \Eprint
  {http://arxiv.org/abs/hep-ph/0308085} {arXiv:hep-ph/0308085 [hep-ph]}
  \BibitemShut {NoStop}%
\bibitem [{\citenamefont {Manninen}\ \emph {et~al.}(2011)\citenamefont
  {Manninen}, \citenamefont {Bratkovskaya}, \citenamefont {Cassing},\ and\
  \citenamefont {Linnyk}}]{Manninen:2010yf}%
  \BibitemOpen
  \bibfield  {author} {\bibinfo {author} {\bibfnamefont {J.}~\bibnamefont
  {Manninen}}, \bibinfo {author} {\bibfnamefont {E.}~\bibnamefont
  {Bratkovskaya}}, \bibinfo {author} {\bibfnamefont {W.}~\bibnamefont
  {Cassing}}, \ and\ \bibinfo {author} {\bibfnamefont {O.}~\bibnamefont
  {Linnyk}},\ }\href {\doibase 10.1140/epjc/s10052-011-1615-4} {\bibfield
  {journal} {\bibinfo  {journal} {Eur.Phys.J.}\ }\textbf {\bibinfo {volume}
  {C71}},\ \bibinfo {pages} {1615} (\bibinfo {year} {2011})},\ \Eprint
  {http://arxiv.org/abs/1005.0500} {arXiv:1005.0500 [nucl-th]} \BibitemShut
  {NoStop}%
\bibitem [{\citenamefont {Staig}\ and\ \citenamefont
  {Shuryak}(2010)}]{Staig:2010by}%
  \BibitemOpen
  \bibfield  {author} {\bibinfo {author} {\bibfnamefont {P.}~\bibnamefont
  {Staig}}\ and\ \bibinfo {author} {\bibfnamefont {E.}~\bibnamefont
  {Shuryak}},\ }\href@noop {} {\  (\bibinfo {year} {2010})},\ \Eprint
  {http://arxiv.org/abs/1005.3531} {arXiv:1005.3531 [nucl-th]} \BibitemShut
  {NoStop}%
\bibitem [{\citenamefont {Linnyk}\ \emph {et~al.}(2012)\citenamefont {Linnyk},
  \citenamefont {Cassing}, \citenamefont {Manninen}, \citenamefont
  {Bratkovskaya},\ and\ \citenamefont {Ko}}]{Linnyk:2011vx}%
  \BibitemOpen
  \bibfield  {author} {\bibinfo {author} {\bibfnamefont {O.}~\bibnamefont
  {Linnyk}}, \bibinfo {author} {\bibfnamefont {W.}~\bibnamefont {Cassing}},
  \bibinfo {author} {\bibfnamefont {J.}~\bibnamefont {Manninen}}, \bibinfo
  {author} {\bibfnamefont {E.}~\bibnamefont {Bratkovskaya}}, \ and\ \bibinfo
  {author} {\bibfnamefont {C.}~\bibnamefont {Ko}},\ }\href {\doibase
  10.1103/PhysRevC.85.024910} {\bibfield  {journal} {\bibinfo  {journal}
  {Phys.Rev.}\ }\textbf {\bibinfo {volume} {C85}},\ \bibinfo {pages} {024910}
  (\bibinfo {year} {2012})},\ \Eprint {http://arxiv.org/abs/1111.2975}
  {arXiv:1111.2975 [nucl-th]} \BibitemShut {NoStop}%
\bibitem [{\citenamefont {Linnyk}\ \emph
  {et~al.}(2013{\natexlab{a}})\citenamefont {Linnyk}, \citenamefont {Cassing},
  \citenamefont {Manninen}, \citenamefont {Bratkovskaya}, \citenamefont
  {Gossiaux} \emph {et~al.}}]{Linnyk:2012pu}%
  \BibitemOpen
  \bibfield  {author} {\bibinfo {author} {\bibfnamefont {O.}~\bibnamefont
  {Linnyk}}, \bibinfo {author} {\bibfnamefont {W.}~\bibnamefont {Cassing}},
  \bibinfo {author} {\bibfnamefont {J.}~\bibnamefont {Manninen}}, \bibinfo
  {author} {\bibfnamefont {E.}~\bibnamefont {Bratkovskaya}}, \bibinfo {author}
  {\bibfnamefont {P.}~\bibnamefont {Gossiaux}},  \emph {et~al.},\ }\href
  {\doibase 10.1103/PhysRevC.87.014905} {\bibfield  {journal} {\bibinfo
  {journal} {Phys.Rev.}\ }\textbf {\bibinfo {volume} {C87}},\ \bibinfo {pages}
  {014905} (\bibinfo {year} {2013}{\natexlab{a}})},\ \Eprint
  {http://arxiv.org/abs/1208.1279} {arXiv:1208.1279 [nucl-th]} \BibitemShut
  {NoStop}%
\bibitem [{\citenamefont {Rapp}(2013)}]{Rapp:2013nxa}%
  \BibitemOpen
  \bibfield  {author} {\bibinfo {author} {\bibfnamefont {R.}~\bibnamefont
  {Rapp}},\ }\href {\doibase 10.1155/2013/148253} {\bibfield  {journal}
  {\bibinfo  {journal} {Adv.High Energy Phys.}\ }\textbf {\bibinfo {volume}
  {2013}},\ \bibinfo {pages} {148253} (\bibinfo {year} {2013})},\ \Eprint
  {http://arxiv.org/abs/1304.2309} {arXiv:1304.2309 [hep-ph]} \BibitemShut
  {NoStop}%
\bibitem [{\citenamefont {Hohler}\ and\ \citenamefont
  {Rapp}(2014)}]{Hohler:2013eba}%
  \BibitemOpen
  \bibfield  {author} {\bibinfo {author} {\bibfnamefont {P.~M.}\ \bibnamefont
  {Hohler}}\ and\ \bibinfo {author} {\bibfnamefont {R.}~\bibnamefont {Rapp}},\
  }\href {\doibase 10.1016/j.physletb.2014.02.021} {\bibfield  {journal}
  {\bibinfo  {journal} {Phys.Lett.}\ }\textbf {\bibinfo {volume} {B731}},\
  \bibinfo {pages} {103} (\bibinfo {year} {2014})},\ \Eprint
  {http://arxiv.org/abs/1311.2921} {arXiv:1311.2921 [hep-ph]} \BibitemShut
  {NoStop}%
\bibitem [{\citenamefont {Lee}\ and\ \citenamefont
  {Zahed}(2014)}]{Lee:2014pwa}%
  \BibitemOpen
  \bibfield  {author} {\bibinfo {author} {\bibfnamefont {C.-H.}\ \bibnamefont
  {Lee}}\ and\ \bibinfo {author} {\bibfnamefont {I.}~\bibnamefont {Zahed}},\
  }\href@noop {} {\  (\bibinfo {year} {2014})},\ \Eprint
  {http://arxiv.org/abs/1403.1632} {arXiv:1403.1632 [hep-ph]} \BibitemShut
  {NoStop}%
\bibitem [{\citenamefont {Vujanovic}\ \emph {et~al.}(2013)\citenamefont
  {Vujanovic}, \citenamefont {Young}, \citenamefont {Schenke}, \citenamefont
  {Rapp}, \citenamefont {Jeon},\ and\ \citenamefont
  {Gale}}]{Vujanovic:2013jpa}%
  \BibitemOpen
  \bibfield  {author} {\bibinfo {author} {\bibfnamefont {G.}~\bibnamefont
  {Vujanovic}}, \bibinfo {author} {\bibfnamefont {C.}~\bibnamefont {Young}},
  \bibinfo {author} {\bibfnamefont {B.}~\bibnamefont {Schenke}}, \bibinfo
  {author} {\bibfnamefont {R.}~\bibnamefont {Rapp}}, \bibinfo {author}
  {\bibfnamefont {S.}~\bibnamefont {Jeon}}, \ and\ \bibinfo {author}
  {\bibfnamefont {C.}~\bibnamefont {Gale}},\ }\href@noop {} {\  (\bibinfo
  {year} {2013})},\ \Eprint {http://arxiv.org/abs/1312.0676} {arXiv:1312.0676
  [nucl-th]} \BibitemShut {NoStop}%
\bibitem [{\citenamefont {Chatterjee}\ \emph {et~al.}(2006)\citenamefont
  {Chatterjee}, \citenamefont {Frodermann}, \citenamefont {Heinz},\ and\
  \citenamefont {Srivastava}}]{Chatterjee:2005de}%
  \BibitemOpen
  \bibfield  {author} {\bibinfo {author} {\bibfnamefont {R.}~\bibnamefont
  {Chatterjee}}, \bibinfo {author} {\bibfnamefont {E.~S.}\ \bibnamefont
  {Frodermann}}, \bibinfo {author} {\bibfnamefont {U.~W.}\ \bibnamefont
  {Heinz}}, \ and\ \bibinfo {author} {\bibfnamefont {D.~K.}\ \bibnamefont
  {Srivastava}},\ }\href {\doibase 10.1103/PhysRevLett.96.202302} {\bibfield
  {journal} {\bibinfo  {journal} {Phys.Rev.Lett.}\ }\textbf {\bibinfo {volume}
  {96}},\ \bibinfo {pages} {202302} (\bibinfo {year} {2006})},\ \Eprint
  {http://arxiv.org/abs/nucl-th/0511079} {arXiv:nucl-th/0511079 [nucl-th]}
  \BibitemShut {NoStop}%
\bibitem [{\citenamefont {Bratkovskaya}\ \emph {et~al.}(2008)\citenamefont
  {Bratkovskaya}, \citenamefont {Kiselev},\ and\ \citenamefont
  {Sharkov}}]{Bratkovskaya:2008iq}%
  \BibitemOpen
  \bibfield  {author} {\bibinfo {author} {\bibfnamefont {E.}~\bibnamefont
  {Bratkovskaya}}, \bibinfo {author} {\bibfnamefont {S.}~\bibnamefont
  {Kiselev}}, \ and\ \bibinfo {author} {\bibfnamefont {G.}~\bibnamefont
  {Sharkov}},\ }\href {\doibase 10.1103/PhysRevC.78.034905} {\bibfield
  {journal} {\bibinfo  {journal} {Phys.Rev.}\ }\textbf {\bibinfo {volume}
  {C78}},\ \bibinfo {pages} {034905} (\bibinfo {year} {2008})},\ \Eprint
  {http://arxiv.org/abs/0806.3465} {arXiv:0806.3465 [nucl-th]} \BibitemShut
  {NoStop}%
\bibitem [{\citenamefont {van Hees}\ \emph {et~al.}(2011)\citenamefont {van
  Hees}, \citenamefont {Gale},\ and\ \citenamefont {Rapp}}]{vanHees:2011vb}%
  \BibitemOpen
  \bibfield  {author} {\bibinfo {author} {\bibfnamefont {H.}~\bibnamefont {van
  Hees}}, \bibinfo {author} {\bibfnamefont {C.}~\bibnamefont {Gale}}, \ and\
  \bibinfo {author} {\bibfnamefont {R.}~\bibnamefont {Rapp}},\ }\href {\doibase
  10.1103/PhysRevC.84.054906} {\bibfield  {journal} {\bibinfo  {journal}
  {Phys.Rev.}\ }\textbf {\bibinfo {volume} {C84}},\ \bibinfo {pages} {054906}
  (\bibinfo {year} {2011})},\ \Eprint {http://arxiv.org/abs/1108.2131}
  {arXiv:1108.2131 [hep-ph]} \BibitemShut {NoStop}%
\bibitem [{\citenamefont {Basar}\ \emph {et~al.}(2012)\citenamefont {Basar},
  \citenamefont {Kharzeev},\ and\ \citenamefont {Skokov}}]{Basar:2012bp}%
  \BibitemOpen
  \bibfield  {author} {\bibinfo {author} {\bibfnamefont {G.}~\bibnamefont
  {Basar}}, \bibinfo {author} {\bibfnamefont {D.}~\bibnamefont {Kharzeev}}, \
  and\ \bibinfo {author} {\bibfnamefont {V.}~\bibnamefont {Skokov}},\ }\href
  {\doibase 10.1103/PhysRevLett.109.202303} {\bibfield  {journal} {\bibinfo
  {journal} {Phys.Rev.Lett.}\ }\textbf {\bibinfo {volume} {109}},\ \bibinfo
  {pages} {202303} (\bibinfo {year} {2012})},\ \Eprint
  {http://arxiv.org/abs/1206.1334} {arXiv:1206.1334 [hep-ph]} \BibitemShut
  {NoStop}%
\bibitem [{\citenamefont {Bzdak}\ and\ \citenamefont
  {Skokov}(2013)}]{Bzdak:2012fr}%
  \BibitemOpen
  \bibfield  {author} {\bibinfo {author} {\bibfnamefont {A.}~\bibnamefont
  {Bzdak}}\ and\ \bibinfo {author} {\bibfnamefont {V.}~\bibnamefont {Skokov}},\
  }\href {\doibase 10.1103/PhysRevLett.110.192301} {\bibfield  {journal}
  {\bibinfo  {journal} {Phys.Rev.Lett.}\ }\textbf {\bibinfo {volume} {110}},\
  \bibinfo {pages} {192301} (\bibinfo {year} {2013})},\ \Eprint
  {http://arxiv.org/abs/1208.5502} {arXiv:1208.5502 [hep-ph]} \BibitemShut
  {NoStop}%
\bibitem [{\citenamefont {Fukushima}\ and\ \citenamefont
  {Mameda}(2012)}]{Fukushima:2012fg}%
  \BibitemOpen
  \bibfield  {author} {\bibinfo {author} {\bibfnamefont {K.}~\bibnamefont
  {Fukushima}}\ and\ \bibinfo {author} {\bibfnamefont {K.}~\bibnamefont
  {Mameda}},\ }\href {\doibase 10.1103/PhysRevD.86.071501} {\bibfield
  {journal} {\bibinfo  {journal} {Phys.Rev.}\ }\textbf {\bibinfo {volume}
  {D86}},\ \bibinfo {pages} {071501} (\bibinfo {year} {2012})},\ \Eprint
  {http://arxiv.org/abs/1206.3128} {arXiv:1206.3128 [hep-ph]} \BibitemShut
  {NoStop}%
\bibitem [{\citenamefont {Liu}\ and\ \citenamefont {Liu}(2014)}]{Liu:2012ax}%
  \BibitemOpen
  \bibfield  {author} {\bibinfo {author} {\bibfnamefont {F.-M.}\ \bibnamefont
  {Liu}}\ and\ \bibinfo {author} {\bibfnamefont {S.-X.}\ \bibnamefont {Liu}},\
  }\href {\doibase 10.1103/PhysRevC.89.034906} {\bibfield  {journal} {\bibinfo
  {journal} {Phys.Rev.}\ }\textbf {\bibinfo {volume} {C89}},\ \bibinfo {pages}
  {034906} (\bibinfo {year} {2014})},\ \Eprint {http://arxiv.org/abs/1212.6587}
  {arXiv:1212.6587 [nucl-th]} \BibitemShut {NoStop}%
\bibitem [{\citenamefont {Shen}\ \emph {et~al.}(2013)\citenamefont {Shen},
  \citenamefont {Heinz}, \citenamefont {Paquet}, \citenamefont {Kozlov},\ and\
  \citenamefont {Gale}}]{Shen:2013cca}%
  \BibitemOpen
  \bibfield  {author} {\bibinfo {author} {\bibfnamefont {C.}~\bibnamefont
  {Shen}}, \bibinfo {author} {\bibfnamefont {U.~W.}\ \bibnamefont {Heinz}},
  \bibinfo {author} {\bibfnamefont {J.-F.}\ \bibnamefont {Paquet}}, \bibinfo
  {author} {\bibfnamefont {I.}~\bibnamefont {Kozlov}}, \ and\ \bibinfo {author}
  {\bibfnamefont {C.}~\bibnamefont {Gale}},\ }\href@noop {} {\  (\bibinfo
  {year} {2013})},\ \Eprint {http://arxiv.org/abs/1308.2111} {arXiv:1308.2111
  [nucl-th]} \BibitemShut {NoStop}%
\bibitem [{\citenamefont {Shen}\ \emph {et~al.}(2014)\citenamefont {Shen},
  \citenamefont {Heinz}, \citenamefont {Paquet},\ and\ \citenamefont
  {Gale}}]{Shen:2013vja}%
  \BibitemOpen
  \bibfield  {author} {\bibinfo {author} {\bibfnamefont {C.}~\bibnamefont
  {Shen}}, \bibinfo {author} {\bibfnamefont {U.~W.}\ \bibnamefont {Heinz}},
  \bibinfo {author} {\bibfnamefont {J.-F.}\ \bibnamefont {Paquet}}, \ and\
  \bibinfo {author} {\bibfnamefont {C.}~\bibnamefont {Gale}},\ }\href {\doibase
  10.1103/PhysRevC.89.044910} {\bibfield  {journal} {\bibinfo  {journal}
  {Phys.Rev.}\ }\textbf {\bibinfo {volume} {C89}},\ \bibinfo {pages} {044910}
  (\bibinfo {year} {2014})},\ \Eprint {http://arxiv.org/abs/1308.2440}
  {arXiv:1308.2440 [nucl-th]} \BibitemShut {NoStop}%
\bibitem [{\citenamefont {Linnyk}\ \emph
  {et~al.}(2013{\natexlab{b}})\citenamefont {Linnyk}, \citenamefont
  {Konchakovski}, \citenamefont {Cassing},\ and\ \citenamefont
  {Bratkovskaya}}]{Linnyk:2013hta}%
  \BibitemOpen
  \bibfield  {author} {\bibinfo {author} {\bibfnamefont {O.}~\bibnamefont
  {Linnyk}}, \bibinfo {author} {\bibfnamefont {V.}~\bibnamefont
  {Konchakovski}}, \bibinfo {author} {\bibfnamefont {W.}~\bibnamefont
  {Cassing}}, \ and\ \bibinfo {author} {\bibfnamefont {E.}~\bibnamefont
  {Bratkovskaya}},\ }\href {\doibase 10.1103/PhysRevC.88.034904} {\bibfield
  {journal} {\bibinfo  {journal} {Phys.Rev.}\ }\textbf {\bibinfo {volume}
  {C88}},\ \bibinfo {pages} {034904} (\bibinfo {year} {2013}{\natexlab{b}})},\
  \Eprint {http://arxiv.org/abs/1304.7030} {arXiv:1304.7030 [nucl-th]}
  \BibitemShut {NoStop}%
\bibitem [{\citenamefont {Linnyk}\ \emph {et~al.}(2014)\citenamefont {Linnyk},
  \citenamefont {Cassing},\ and\ \citenamefont
  {Bratkovskaya}}]{Linnyk:2013wma}%
  \BibitemOpen
  \bibfield  {author} {\bibinfo {author} {\bibfnamefont {O.}~\bibnamefont
  {Linnyk}}, \bibinfo {author} {\bibfnamefont {W.}~\bibnamefont {Cassing}}, \
  and\ \bibinfo {author} {\bibfnamefont {E.}~\bibnamefont {Bratkovskaya}},\
  }\href {\doibase 10.1103/PhysRevC.89.034908} {\bibfield  {journal} {\bibinfo
  {journal} {Phys.Rev.}\ }\textbf {\bibinfo {volume} {C89}},\ \bibinfo {pages}
  {034908} (\bibinfo {year} {2014})},\ \Eprint {http://arxiv.org/abs/1311.0279}
  {arXiv:1311.0279 [nucl-th]} \BibitemShut {NoStop}%
\bibitem [{\citenamefont {Muller}\ \emph {et~al.}(2014)\citenamefont {Muller},
  \citenamefont {Wu},\ and\ \citenamefont {Yang}}]{Muller:2013ila}%
  \BibitemOpen
  \bibfield  {author} {\bibinfo {author} {\bibfnamefont {B.}~\bibnamefont
  {Muller}}, \bibinfo {author} {\bibfnamefont {S.-Y.}\ \bibnamefont {Wu}}, \
  and\ \bibinfo {author} {\bibfnamefont {D.-L.}\ \bibnamefont {Yang}},\ }\href
  {\doibase 10.1103/PhysRevD.89.026013} {\bibfield  {journal} {\bibinfo
  {journal} {Phys.Rev.}\ }\textbf {\bibinfo {volume} {D89}},\ \bibinfo {pages}
  {026013} (\bibinfo {year} {2014})},\ \Eprint {http://arxiv.org/abs/1308.6568}
  {arXiv:1308.6568 [hep-th]} \BibitemShut {NoStop}%
\bibitem [{\citenamefont {Basar}\ \emph {et~al.}(2014)\citenamefont {Basar},
  \citenamefont {Kharzeev},\ and\ \citenamefont {Shuryak}}]{Basar:2014swa}%
  \BibitemOpen
  \bibfield  {author} {\bibinfo {author} {\bibfnamefont {G.}~\bibnamefont
  {Basar}}, \bibinfo {author} {\bibfnamefont {D.~E.}\ \bibnamefont {Kharzeev}},
  \ and\ \bibinfo {author} {\bibfnamefont {E.~V.}\ \bibnamefont {Shuryak}},\
  }\href@noop {} {\  (\bibinfo {year} {2014})},\ \Eprint
  {http://arxiv.org/abs/1402.2286} {arXiv:1402.2286 [hep-ph]} \BibitemShut
  {NoStop}%
\bibitem [{\citenamefont {van Hees}\ \emph {et~al.}(2014)\citenamefont {van
  Hees}, \citenamefont {He},\ and\ \citenamefont {Rapp}}]{vanHees:2014ida}%
  \BibitemOpen
  \bibfield  {author} {\bibinfo {author} {\bibfnamefont {H.}~\bibnamefont {van
  Hees}}, \bibinfo {author} {\bibfnamefont {M.}~\bibnamefont {He}}, \ and\
  \bibinfo {author} {\bibfnamefont {R.}~\bibnamefont {Rapp}},\ }\href@noop {}
  {\  (\bibinfo {year} {2014})},\ \Eprint {http://arxiv.org/abs/1404.2846}
  {arXiv:1404.2846 [nucl-th]} \BibitemShut {NoStop}%
\bibitem [{\citenamefont {Monnai}(2014)}]{Monnai:2014kqa}%
  \BibitemOpen
  \bibfield  {author} {\bibinfo {author} {\bibfnamefont {A.}~\bibnamefont
  {Monnai}},\ }\href@noop {} {\  (\bibinfo {year} {2014})},\ \Eprint
  {http://arxiv.org/abs/1403.4225} {arXiv:1403.4225 [nucl-th]} \BibitemShut
  {NoStop}%
\bibitem [{\citenamefont {McLerran}\ and\ \citenamefont
  {Schenke}(2014)}]{McLerran:2014hza}%
  \BibitemOpen
  \bibfield  {author} {\bibinfo {author} {\bibfnamefont {L.}~\bibnamefont
  {McLerran}}\ and\ \bibinfo {author} {\bibfnamefont {B.}~\bibnamefont
  {Schenke}},\ }\href@noop {} {\  (\bibinfo {year} {2014})},\ \Eprint
  {http://arxiv.org/abs/1403.7462} {arXiv:1403.7462 [hep-ph]} \BibitemShut
  {NoStop}%
\bibitem [{\citenamefont {Dion}\ \emph {et~al.}(2011)\citenamefont {Dion},
  \citenamefont {Paquet}, \citenamefont {Schenke}, \citenamefont {Young},
  \citenamefont {Jeon} \emph {et~al.}}]{Dion:2011pp}%
  \BibitemOpen
  \bibfield  {author} {\bibinfo {author} {\bibfnamefont {M.}~\bibnamefont
  {Dion}}, \bibinfo {author} {\bibfnamefont {J.-F.}\ \bibnamefont {Paquet}},
  \bibinfo {author} {\bibfnamefont {B.}~\bibnamefont {Schenke}}, \bibinfo
  {author} {\bibfnamefont {C.}~\bibnamefont {Young}}, \bibinfo {author}
  {\bibfnamefont {S.}~\bibnamefont {Jeon}},  \emph {et~al.},\ }\href {\doibase
  10.1103/PhysRevC.84.064901} {\bibfield  {journal} {\bibinfo  {journal}
  {Phys.Rev.}\ }\textbf {\bibinfo {volume} {C84}},\ \bibinfo {pages} {064901}
  (\bibinfo {year} {2011})},\ \Eprint {http://arxiv.org/abs/1109.4405}
  {arXiv:1109.4405 [hep-ph]} \BibitemShut {NoStop}%
\bibitem [{\citenamefont {Klasen}\ \emph {et~al.}(2013)\citenamefont {Klasen},
  \citenamefont {Klein-Boesing}, \citenamefont {König},\ and\ \citenamefont
  {Wessels}}]{Klasen:2013mga}%
  \BibitemOpen
  \bibfield  {author} {\bibinfo {author} {\bibfnamefont {M.}~\bibnamefont
  {Klasen}}, \bibinfo {author} {\bibfnamefont {C.}~\bibnamefont
  {Klein-Boesing}}, \bibinfo {author} {\bibfnamefont {F.}~\bibnamefont
  {König}}, \ and\ \bibinfo {author} {\bibfnamefont {J.}~\bibnamefont
  {Wessels}},\ }\href {\doibase 10.1007/JHEP10(2013)119} {\bibfield  {journal}
  {\bibinfo  {journal} {JHEP}\ }\textbf {\bibinfo {volume} {1310}},\ \bibinfo
  {pages} {119} (\bibinfo {year} {2013})},\ \Eprint
  {http://arxiv.org/abs/1307.7034} {arXiv:1307.7034} \BibitemShut {NoStop}%
\bibitem [{\citenamefont {Klasen}\ and\ \citenamefont
  {Konig}(2014)}]{Klasen:2014xfa}%
  \BibitemOpen
  \bibfield  {author} {\bibinfo {author} {\bibfnamefont {M.}~\bibnamefont
  {Klasen}}\ and\ \bibinfo {author} {\bibfnamefont {F.}~\bibnamefont {Konig}},\
  }\href@noop {} {\  (\bibinfo {year} {2014})},\ \Eprint
  {http://arxiv.org/abs/1403.2290} {arXiv:1403.2290 [hep-ph]} \BibitemShut
  {NoStop}%
\bibitem [{\citenamefont {Arleo}(2006)}]{Arleo:2006xb}%
  \BibitemOpen
  \bibfield  {author} {\bibinfo {author} {\bibfnamefont {F.}~\bibnamefont
  {Arleo}},\ }\href {\doibase 10.1088/1126-6708/2006/09/015} {\bibfield
  {journal} {\bibinfo  {journal} {JHEP}\ }\textbf {\bibinfo {volume} {0609}},\
  \bibinfo {pages} {015} (\bibinfo {year} {2006})},\ \Eprint
  {http://arxiv.org/abs/hep-ph/0601075} {arXiv:hep-ph/0601075 [hep-ph]}
  \BibitemShut {NoStop}%
\bibitem [{\citenamefont {Arleo}(2009)}]{Arleo:2008dn}%
  \BibitemOpen
  \bibfield  {author} {\bibinfo {author} {\bibfnamefont {F.}~\bibnamefont
  {Arleo}},\ }\href {\doibase 10.1140/epjc/s10052-009-0871-z} {\bibfield
  {journal} {\bibinfo  {journal} {Eur.Phys.J.}\ }\textbf {\bibinfo {volume}
  {C61}},\ \bibinfo {pages} {603} (\bibinfo {year} {2009})},\ \Eprint
  {http://arxiv.org/abs/0810.1193} {arXiv:0810.1193 [hep-ph]} \BibitemShut
  {NoStop}%
\bibitem [{\citenamefont {Arleo}\ \emph {et~al.}(2011)\citenamefont {Arleo},
  \citenamefont {Eskola}, \citenamefont {Paukkunen},\ and\ \citenamefont
  {Salgado}}]{Arleo:2011gc}%
  \BibitemOpen
  \bibfield  {author} {\bibinfo {author} {\bibfnamefont {F.}~\bibnamefont
  {Arleo}}, \bibinfo {author} {\bibfnamefont {K.~J.}\ \bibnamefont {Eskola}},
  \bibinfo {author} {\bibfnamefont {H.}~\bibnamefont {Paukkunen}}, \ and\
  \bibinfo {author} {\bibfnamefont {C.~A.}\ \bibnamefont {Salgado}},\ }\href
  {\doibase 10.1007/JHEP04(2011)055} {\bibfield  {journal} {\bibinfo  {journal}
  {JHEP}\ }\textbf {\bibinfo {volume} {1104}},\ \bibinfo {pages} {055}
  (\bibinfo {year} {2011})},\ \Eprint {http://arxiv.org/abs/1103.1471}
  {arXiv:1103.1471 [hep-ph]} \BibitemShut {NoStop}%
\bibitem [{\citenamefont {Adare}\ \emph
  {et~al.}(2012{\natexlab{a}})\citenamefont {Adare} \emph
  {et~al.}}]{Adare:2011zr}%
  \BibitemOpen
  \bibfield  {author} {\bibinfo {author} {\bibfnamefont {A.}~\bibnamefont
  {Adare}} \emph {et~al.} (\bibinfo {collaboration} {PHENIX Collaboration}),\
  }\href {\doibase 10.1103/PhysRevLett.109.122302} {\bibfield  {journal}
  {\bibinfo  {journal} {Phys.Rev.Lett.}\ }\textbf {\bibinfo {volume} {109}},\
  \bibinfo {pages} {122302} (\bibinfo {year} {2012}{\natexlab{a}})},\ \Eprint
  {http://arxiv.org/abs/1105.4126} {arXiv:1105.4126 [nucl-ex]} \BibitemShut
  {NoStop}%
\bibitem [{\citenamefont {Lohner}(2013)}]{Lohner:2012ct}%
  \BibitemOpen
  \bibfield  {author} {\bibinfo {author} {\bibfnamefont {D.}~\bibnamefont
  {Lohner}} (\bibinfo {collaboration} {ALICE}),\ }\href {\doibase
  10.1088/1742-6596/446/1/012028} {\bibfield  {journal} {\bibinfo  {journal}
  {J.Phys.Conf.Ser.}\ }\textbf {\bibinfo {volume} {446}},\ \bibinfo {pages}
  {012028} (\bibinfo {year} {2013})},\ \Eprint {http://arxiv.org/abs/1212.3995}
  {arXiv:1212.3995 [hep-ex]} \BibitemShut {NoStop}%
\bibitem [{\citenamefont {Sakaguchi}(2014)}]{Sakaguchi:2014ewa}%
  \BibitemOpen
  \bibfield  {author} {\bibinfo {author} {\bibfnamefont {T.}~\bibnamefont
  {Sakaguchi}},\ }\href@noop {} {\  (\bibinfo {year} {2014})},\ \Eprint
  {http://arxiv.org/abs/1401.2481} {arXiv:1401.2481 [nucl-ex]} \BibitemShut
  {NoStop}%
\bibitem [{\citenamefont {Hidaka}\ \emph {et~al.}(2014)\citenamefont {Hidaka},
  \citenamefont {Lin}, \citenamefont {Pisarski},\ and\ \citenamefont
  {Satow}}]{Future}%
  \BibitemOpen
  \bibfield  {author} {\bibinfo {author} {\bibfnamefont {Y.}~\bibnamefont
  {Hidaka}}, \bibinfo {author} {\bibfnamefont {S.}~\bibnamefont {Lin}},
  \bibinfo {author} {\bibfnamefont {R.~D.}\ \bibnamefont {Pisarski}}, \ and\
  \bibinfo {author} {\bibfnamefont {D.}~\bibnamefont {Satow}},\ }\href@noop {}
  {\bibfield  {journal} {\bibinfo  {journal} {in preparation}\ } (\bibinfo
  {year} {2014})}\BibitemShut {NoStop}%
\bibitem [{\citenamefont {Gava}\ and\ \citenamefont
  {Jengo}(1981)}]{Gava:1981qd}%
  \BibitemOpen
  \bibfield  {author} {\bibinfo {author} {\bibfnamefont {E.}~\bibnamefont
  {Gava}}\ and\ \bibinfo {author} {\bibfnamefont {R.}~\bibnamefont {Jengo}},\
  }\href {\doibase 10.1016/0370-2693(81)90890-X} {\bibfield  {journal}
  {\bibinfo  {journal} {Phys.Lett.}\ }\textbf {\bibinfo {volume} {B105}},\
  \bibinfo {pages} {285} (\bibinfo {year} {1981})}\BibitemShut {NoStop}%
\bibitem [{\citenamefont {Burnier}\ \emph {et~al.}(2010)\citenamefont
  {Burnier}, \citenamefont {Laine},\ and\ \citenamefont
  {Vepsalainen}}]{Burnier:2009bk}%
  \BibitemOpen
  \bibfield  {author} {\bibinfo {author} {\bibfnamefont {Y.}~\bibnamefont
  {Burnier}}, \bibinfo {author} {\bibfnamefont {M.}~\bibnamefont {Laine}}, \
  and\ \bibinfo {author} {\bibfnamefont {M.}~\bibnamefont {Vepsalainen}},\
  }\href {\doibase 10.1007/JHEP01(2010)054, 10.1007/JHEP01(2013)180} {\bibfield
   {journal} {\bibinfo  {journal} {JHEP}\ }\textbf {\bibinfo {volume} {1001}},\
  \bibinfo {pages} {054} (\bibinfo {year} {2010})},\ \Eprint
  {http://arxiv.org/abs/0911.3480} {arXiv:0911.3480 [hep-ph]} \BibitemShut
  {NoStop}%
\bibitem [{\citenamefont {Brambilla}\ \emph {et~al.}(2010)\citenamefont
  {Brambilla}, \citenamefont {Ghiglieri}, \citenamefont {Petreczky},\ and\
  \citenamefont {Vairo}}]{Brambilla:2010xn}%
  \BibitemOpen
  \bibfield  {author} {\bibinfo {author} {\bibfnamefont {N.}~\bibnamefont
  {Brambilla}}, \bibinfo {author} {\bibfnamefont {J.}~\bibnamefont
  {Ghiglieri}}, \bibinfo {author} {\bibfnamefont {P.}~\bibnamefont
  {Petreczky}}, \ and\ \bibinfo {author} {\bibfnamefont {A.}~\bibnamefont
  {Vairo}},\ }\href {\doibase 10.1103/PhysRevD.82.074019} {\bibfield  {journal}
  {\bibinfo  {journal} {Phys.Rev.}\ }\textbf {\bibinfo {volume} {D82}},\
  \bibinfo {pages} {074019} (\bibinfo {year} {2010})},\ \Eprint
  {http://arxiv.org/abs/1007.5172} {arXiv:1007.5172 [hep-ph]} \BibitemShut
  {NoStop}%
\bibitem [{\citenamefont {Adare}\ \emph
  {et~al.}(2012{\natexlab{b}})\citenamefont {Adare} \emph
  {et~al.}}]{Adare:2012yt}%
  \BibitemOpen
  \bibfield  {author} {\bibinfo {author} {\bibfnamefont {A.}~\bibnamefont
  {Adare}} \emph {et~al.} (\bibinfo {collaboration} {PHENIX Collaboration}),\
  }\href {\doibase 10.1103/PhysRevD.86.072008} {\bibfield  {journal} {\bibinfo
  {journal} {Phys.Rev.}\ }\textbf {\bibinfo {volume} {D86}},\ \bibinfo {pages}
  {072008} (\bibinfo {year} {2012}{\natexlab{b}})},\ \Eprint
  {http://arxiv.org/abs/1205.5533} {arXiv:1205.5533 [hep-ex]} \BibitemShut
  {NoStop}%
\end{thebibliography}%

\end{document}